\documentclass[twocolumn,rmp,showpacs,preprintnumbers,amsmath,amssymb]{revtex4}
\usepackage{graphicx}
\usepackage{latexsym}

\usepackage{amssymb,amsmath}

\newcommand{\stoy}{\ensuremath{SU(2) \x U(1)_Y}}  
\newcommand{\sthtoy}{\ensuremath{SU(3) \x SU(2) \x U(1)_Y}}                                                    
\newcommand{\eqal}[1]{\label{#1} \end{eqnarray}}   

\newcommand{\ra}{\ensuremath{\rightarrow}}

\newcommand{\zpr}{\ensuremath{Z'}}
\newcommand{\mzp}{\ensuremath{M_{Z'}}}

\newcommand{\uprm}{\ensuremath{U(1)'}}

\newcommand{\msb}{\ensuremath{\overline{\rm{MS}}\ }}

\newcommand{\skipblk}[1]{}                                                      
\def\bqa{\begin{eqnarray}}                                                      
\def\eqa{\end{eqnarray}}

\newcommand{\douba}[2]{\ensuremath{                                                  
\left( \begin{array}{c} #1    \\ #2 
        \end{array}\right)}}

\newcommand{\x}{\ensuremath{\times}}

\newcommand{\sinn}{\ensuremath{\sin^2\theta_W\,}}

\newcommand{\beq}{\begin{equation}}                                             
\newcommand{\eeq}{\end{equation}}     
\newcommand{\eeql}[1]{\label{#1} \end{equation}}

\newcommand{\oh}{\ensuremath{\frac{1}{2}}}

\begin{document}

\title{The Physics of Heavy \zpr\ Gauge Bosons}

\author{Paul Langacker}
\affiliation{School of Natural Sciences, Institute for Advanced Study,
Princeton, NJ 08540\\
email: pgl@ias.edu}

\begin{abstract}
The \uprm\ symmetry associated with a possible heavy \zpr\ would have profound implications for
particle physics and cosmology.
The motivations for such particles in various extensions of the standard model, 
possible ranges for their masses
and couplings, and classes of anomaly-free models are discussed. Present limits from electroweak and
collider experiments are briefly surveyed, as are prospects for discovery and diagnostic
study at future colliders. Implications of a \zpr\ are discussed,
including an extended Higgs sector, extended neutralino sector, and solution to the
$\mu$ problem in supersymmetry; exotic fermions needed for anomaly cancellation; possible
flavor changing neutral current effects; neutrino mass; possible \zpr\ mediation of supersymmetry
breaking; and cosmological implications for cold dark matter and electroweak baryogenesis. 

\end{abstract}

\pacs{12.60.Cn, 12.60.Fr, 14.70.Pw}
\maketitle
\tableofcontents

\section{Introduction} \label{sec:int}
Additional \uprm\ gauge symmetries and associated \zpr\ gauge bosons
are one of the best motivated extensions of the standard model (SM).
It is not so much that they solve any problems as the fact that it is more difficult to
reduce the rank of an extended gauge group containing 
the standard model than it is to break the non-abelian factors. As a toy example,
consider the gauge group $G=SU(N)$, with $N-1$  diagonal generators. $G$ can be broken by
the vacuum expectation value (VEV) of a real adjoint Higgs representation $\Phi$,
which can be represented by a Hermitian traceless $N\x N$ matrix
\beq
\Phi = \sum_{i=1}^{N^2-1} \varphi^i L_i,
\eeql{adjointhiggs}
where the $\varphi^i$ are the real components of $\Phi$ and the $L_i$ are the fundamental
($N\x N$) representation matrices. When $\Phi$ acquires a VEV $\langle \Phi \rangle$, 
$SU(N)$ is broken to
a subgroup associated with those generators which commute with $\langle \Phi \rangle$. 
Without loss of generality, $\langle \Phi \rangle$ can be  
diagonalized by an $SU(N)$ transformation, so that  the $N-1$
diagonal generators remain unbroken. In special cases some of these may be embedded
in unbroken $SU(K)$ subgroups (when $K$ diagonal elements are equal), but 
the unbroken subgroup always contains at least $U(1)^{N-1}$. 

Soon after the proposal of the electroweak \stoy\ model there were many suggestions for
extended or alternative electroweak gauge theories, some of which
involved additional \uprm\ factors. 
\footnote{
Some examples
include~\cite{Fayet:1977yc,Georgi:1977wk,Mohapatra:1977ew,Barger:1978rj,Davidson:1978pm,Deshpande:1979df,Kim:1979hb,deGroot:1979fi,Rizzo:1979wt,Fayet:1980ss,Barger:1980ix,Masiero:1980dd,Barr:1980fg,delAguila:1981vq,Rizzo:1981dm,Li:1982gc,Barger:1982yr,Barger:1982bj,Barr:1983vu}. 
More complete lists of early references can be found 
in~\cite{Robinett:1981yz,Robinett:1982tq,Langacker:1984dc,Hewett:1988xc}. Previous reviews include~\cite{Hewett:1988xc,delAguila:1993px,Cvetic:1995zs,Cvetic:1997wu,Leike:1998wr,Rizzo:2006nw}.}

An especially compelling motivation came
from the development of grand unified theories larger than the original $SU(5)$ model~\cite{Georgi:1974sy},
such as those based on $SO(10)$ or $E_6$ (See, e.g.,  \cite{Robinett:1981yz,Robinett:1982tq,Langacker:1984dc}. For reviews, see \cite{Langacker:1980js,Hewett:1988xc}.). These had rank larger than
4 and could break to $G_{SM}\x \uprm\,^n, n\ge 1$, where $G_{SM}=\sthtoy$ is the standard model gauge group. However, in the original (non-supersymmetric) versions there was no particular
reason for the additional \zpr\ masses to be at the electroweak or TeV scale where they
could be directly observed.  Similarly, superstring constructions often involve large gauge symmetries
which break to $G_{SM}\x \uprm\,^n$ in the effective four-dimensional theory~\cite{Cvetic:1995rj}, where
some of the \uprm\ are non-anomalous.
 In both string theories and  in supersymmetric versions of
grand unification
with extra \uprm~s  below the string or GUT scale, 
both the \uprm\ and the \stoy\ breaking scales are generally tied to the
soft supersymmetry breaking scale~\cite{Cvetic:1995rj,Cvetic:1996mf,Cvetic:1997wu}. Therefore, if supersymmetry is observed at the LHC
there is a strong motivation that a string or GUT induced  \zpr\ would also have a mass at an observable scale. 
(An exception to this is when
the \uprm\ breaking occurs along a flat 
direction.)

In recent years many TeV scale extensions to the SM have been proposed in addition to supersymmetry,
often with the motivation of resolving the fine tuning associated with the quadratic divergence in the Higgs mass. These include various forms of dynamical symmetry breaking~\cite{Hill:2002ap,Chivukula:2002ry,Chivukula:2003wj} and little Higgs models~\cite{ArkaniHamed:2001nc,Han:2003wu,Han:2005ru,Perelstein:2005ka}, which typically involve extended gauge structures, often
including new \zpr\ gauge bosons at the TeV scale.   Some versions
of theories  with large extra dimensions 
allow the standard model gauge bosons to propagate
freely in the extra dimensions, implying Kaluza-Klein excitations (see, e.g.,~\cite{Antoniadis:1990ew,Masip:1999mk,Casalbuoni:1999ns,Cheung:2001mq,Barbieri:2004qk,Appelquist:2000nn,Cheng:2002ab,Appelquist:2002wb,Gogoladze:2006br,Delgado:1999sv}) of the $Z$ and
other standard model gauge bosons, with effective masses of order $R^{-1}\sim 2 {\ \rm TeV }\x
(10^{-17} {\rm  cm }/R)$, where $R$ is the scale of the extra dimension. Such excitations can also
occur in Randall-Sundrum models~\cite{Randall:1999ee} (see, e.g.,~\cite{Hewett:2002fe,Carena:2003fx,Agashe:2003zs,Agashe:2007ki}).

Other motivations for
new \zpr\ bosons, e.g., associated with (approximately) hidden sectors of nature, 
are detailed in Sections \ref{sec:models} and \ref{sec:implications}. Extensions of the SM
may also involve new TeV scale charged $W$ bosons (see, e.g.,~\cite{Rizzo:2007xs}), which could couple either to
left or right handed currents, but the focus of this article
will be on \zpr~s.

The experimental discovery of a new \zpr\ would be exciting, but the implications would be
much greater than just the existence of a new vector boson. Breaking the \uprm\ symmetry would require an extended
Higgs (and neutralino) sector, with significant consequences for collider physics and cosmology
(direct searches, the $\mu$ problem, dark matter, electroweak  baryogenesis). Anomaly cancellation usually requires
the existence of new exotic particles that are vectorlike with respect to the standard model but
chiral under \uprm, with several possibilities for their decay characteristics. 
The expanded Higgs and exotic sectors can modify or maintain the approximate gauge
coupling unification of the minimal supersymmetric standard model
(MSSM). In some constructions (especially string derived) the \uprm\ charges are
family nonuniversal, which can lead to flavor changing neutral current (FCNC) effects, e.g., in
rare $B$ decays. Finally, the decays of a heavy \zpr\ may be a useful production mechanism for exotics
and superpartners. The constraints from the \uprm\ symmetry can significantly alter the
theoretical possibilities for neutrino mass. Finally, \uprm\ interactions can couple to a hidden sector,
possibly playing a role in supersymmetry breaking or mediation.

Section~\ref{sec:basic} of this review discusses basic issues, such as the \zpr\ interactions and properties, \uprm\ breaking, anomalies, and ordinary and kinetic mixing between $Z$ and \zpr. Section~\ref{sec:models}
surveys the large range of models that have been proposed, including the \uprm-breaking scale;
GUT-inspired models; sets of exotics and charges constructed to avoid anomalies;
and more exotic possibilities such as ultra-weak coupling, low mass, hidden sector,
leptophobic, intermediate scale, sequential, family nonuniversal, and anomalous \uprm\
models.
Section~\ref{sec:exp} briefly outlines the existing constraints from precision electroweak and direct
collider searches, as well as prospects for detection and diagnostics
of couplings at future colliders. Finally, Section~\ref{sec:implications} is a survey
of the theoretical, collider, and cosmological implications of a possible \zpr.

\section{Basic Issues}\label{sec:basic}
\subsection{\zpr\ Couplings}
In the standard model the neutral current interactions of the fermions are described by the 
Lagrangian\footnote{We largely follow the formalism and conventions in
\cite{Durkin:1985ev,Langacker:1991pg}.}
\beq
-L_{\rm NC}^{\rm SM}=gJ^\mu_3 W_{3\mu} + g' J^\mu_Y B_\mu=
eJ^\mu_{em} A_\mu+  g_1 J^\mu_1 Z^0_{1 \mu},
\eeql{LSM}
where $g$ and $g'$ are the $SU(2)$ and $U(1)_Y$ gauge couplings,
$W_{3\mu}$ is the (weak eigenstate) gauge boson associated with the third (diagonal) component of $SU(2)$,
and $B_\mu$ is the $U(1)_Y$ boson. The currents in the first form are

\beq \begin{split}
J^\mu_3 &= \sum_i \bar f_i \gamma^\mu[ t_{3i_{L}}P_L+ t_{3i_{R}}P_R]f_i
\\
J^\mu_Y &= \sum_i \bar f_i \gamma^\mu[y_{i_L}P_L+y_{i_R}P_R]f_i,
\end{split} \eeql{weakcurrent}
where $f_i$ is the field of the $i^{th}$ fermion and
$P_{L,R}\equiv (1\mp\gamma^5)/2$ are the  left and right chiral projections.
$ t_{3i_{L}} ( t_{3i_{R}})$ 
is the third component of weak isospin for
the left (right) chiral component of $f_i$. For the known fermions,
$t_{3u_L}=t_{3\nu_L}=+\oh$, $t_{3d_L}=t_{3e^-_L}=-\oh$, and $t_{3i_R}=0$.
The weak hypercharges $y_{i_{L,R}}$ are chosen to yield the correct electric charges,
\beq
t_{3i_L}+y_{i_L}=t_{3i_R}+y_{i_R}=q_i,
\eeql{hypercharge}
where $q_i$ is the electric charge of $f_i$ in units of the positron charge $e>0$.

 Anticipating the spontaneous breaking of \stoy\ to
the electromagnetic subgroup $U(1)_{em}$ (Section~\ref{sec:ssb}), 
the mass eigenstate  neutral gauge bosons in {Eq.~\ref{LSM}} are the (massless) photon field $A_\mu$
and the (massive) $Z^0_{1 \mu}\equiv Z_\mu$, where
\beq \begin{split}
A_\mu&=\sin \theta_W W_{3\mu} + \cos \theta_W B_\mu \\
Z_\mu&=\cos \theta_W W_{3\mu} -\sin \theta_W B_\mu,
\end{split} \eeql{massbosons}
and the weak angle is
$\theta_W\equiv\tan^{-1} (g'/g)$. The new gauge couplings are
$e=g \sin \theta_W$ and 
\beq
g_1^2 \equiv g^2+ g'\,^2=g^2/\cos^2 \theta_W.
\eeql{g1def}
The currents in the new basis are
\beq \begin{split}
J^\mu_{em}&=\sum_i q_i \bar f_i \gamma^\mu f_i \\
J^\mu_1& =  \sum_i \bar f_i \gamma^\mu[\epsilon_L^{1}(i)P_L+\epsilon_R^{1}(i)P_R]f_i,
\end{split} \eeql{masscurrents}
with the chiral couplings
\beq
\epsilon_L^{1}(i) = t_{3i_{L}}-\sinn q_i, \qquad
\epsilon_R^{1}(i) = t_{3i_{R}}-\sinn q_i.
\eeql{SMcouplings}

In the extension  to $SU(2)\x U(1)_Y\x U(1)'\,^n, n \ge 1$, $L_{\rm NC}$ becomes
\beq
-L_{\rm NC}=eJ^\mu_{em} A_\mu+ \sum_{\alpha=1}^{n+1} g_\alpha J^\mu_\alpha Z^0_{\alpha \mu},
\eeql{lagrangian}
where $g_1, Z^0_{1 \mu}$, and $J^\mu_1$ are respectively the gauge coupling, boson, and
current of the standard model. Similarly, $g_\alpha$ and $Z^0_{\alpha \mu}, \alpha=2\cdots n+1$, are
the gauge couplings  and bosons for the additional \uprm~s.  The currents in Eq.~\ref{lagrangian} are
\beq \begin{split}
J^\mu_\alpha&= \sum_i \bar f_i \gamma^\mu[\epsilon_L^{{\alpha}}(i)P_L+\epsilon_R^{{\alpha}}(i)P_R] f_i
 \\ 
&=\oh  \sum_i \bar f_i \gamma^\mu[g_V^{{\alpha}}(i)-g_A^{{\alpha}}(i)\gamma^5] f_i.
\end{split} \eeql{currents}
The chiral couplings $\epsilon_{L,R}^{{\alpha}}(i)$, which may be unequal for a chiral gauge symmetry,
are respectively the $U(1)_\alpha$  charges of the left and right
handed components of  fermion $f_i$,
and
$g_{V,A}^{{\alpha}}(i) =\epsilon_L^{{\alpha}}(i)\pm\epsilon_R^{{\alpha}}(i)$
are the corresponding vector and axial couplings.

Frequently, it is more convenient to instead specify
 the $U(1)_\alpha$ charges of the left chiral components of \emph{both} the 
fermion $f$ {and} the  antifermion (conjugate) $f^c$, denoted 
$Q_{\alpha f}$ and $Q_{\alpha f^c},$
respectively. The two sets of charges are simply related,
\beq
\epsilon_L^{{\alpha}}(f)=Q_{\alpha f}, \qquad \epsilon_R^{{\alpha}}(f)=-Q_{\alpha f^c}.
\eeql{charges}
For example, in the SM one has $Q_{1u}=\oh - \frac{2}{3} \sinn$ and
$Q_{1u^c}=+ \frac{2}{3} \sinn$.

The additional gauge couplings and charges, as well as the gauge boson masses and mixings, are 
extremely model dependent. The gauge couplings and charges are not independent, i.e.,
one can always replace $g_\alpha$ by $\lambda_\alpha g_\alpha$ provided the
charges $Q_\alpha$ are all simultaneously scaled by $1/\lambda_\alpha$. Usually, the
charges are normalized by some convenient convention.

The three and four point gauge interactions of a complex $SU(2)$ scalar multiplet $\phi$  can be read off from
the kinetic term $L_\phi^{kin}= (D^\mu \phi)^\dagger D_\mu \phi$. The diagonal (neutral current) part of
the gauge covariant derivative of an individual field $\phi_i$ is
\beq
D_\mu \phi_i=\left(  \partial_\mu +i e q_i A_\mu+ i\sum_{\alpha=1}^{n+1} g_\alpha Q_{\alpha i}Z^0_{\alpha \mu} \right) \phi_i,
 \eeql{Dscalar}
where $q_i$ and $Q_{\alpha i}$ are respectively the electric and  $U(1)_\alpha$ charges of $\phi_i$. For the SM part,  $t_i=0, \oh, 1, \cdots$ labels the $SU(2)$ representation, $t_{3i}$
is the third component of weak isospin, the weak hypercharge is $y_i = q_i-t_{3i}$,
and $Q_{1i}=t_{3i}-\sinn q_i$. Thus, for the neutral component $\phi^0$ of the Higgs doublet
$\phi= \douba{\phi^+}{\phi^0}$ one has $t_{\phi^0}=- t_{3\phi^0}=y_{\phi^0}=\oh$.

\subsection{Masses and Mass Mixings}\label{sec:ssb}
We assume that  electrically neutral scalar fields $\phi_i$ acquire VEVs, so
$A_\mu$ remains massless, while the  $Z^0_{\alpha \mu}$ fields develop a
mass term $L_Z^{mass}=\oh M^2_{\alpha\beta} Z^0_{\alpha \mu}Z_{\beta}^{0 \mu}$,
where 
\beq
M^2_{\alpha\beta} = 2 g_\alpha g_\beta\sum_i Q_{\alpha i}Q_{\beta i} |\langle \phi_i \rangle|^2.
\eeql{Zmatrix}
$M^2_{11}\equiv M^2_{Z^0}$ would be the (tree-level) $Z$ mass in the SM limit in which the other $Z^0$'s and
their mixing can be ignored.
 If the only Higgs fields are $SU(2)$ doublets (or singlets), as in the SM or the MSSM, then
\beq
M^2_{Z^0}=\oh g_1^2 \sum_i |\langle \phi_i \rangle|^2 =\frac {1}{4} g_1^2\nu^2=\frac{M_W^2}{\cos^2 \theta_W},
\eeql{SMZ)}
where $\nu^2= 2 \sum_i |\langle \phi_i \rangle|^2\sim (\sqrt{2} G_F)^{-1}\sim (246 {\rm \ GeV})^2$ is the 
square of the weak scale and $G_F$ is the Fermi constant. The observed $Z$ mass strongly constrains
either higher-dimensional Higgs VEVs or $Z-Z'$ mixing~\cite{Yao:2006px}, but in principle they could compensate
and should both be considered. Allowing a general Higgs structure, one has
\beq
M^2_{Z^0}=\frac {g_1^2}{4\sqrt{2} G_F\rho_0}=\frac{M_W^2}{\rho_0\cos^2 \theta_W},
\eeql{rhoMZ}
where 
\beq
\rho_0\equiv\frac{\sum_i (t_i^2-t_{3i}^2+t_i) |\langle \phi_i \rangle|^2}{\sum_i 2t_{3i}^2 |\langle \phi_i \rangle|^2} \xrightarrow[\rm doublets, \ singlets]{} 1.
\eeql{rhopar}

Diagonalizing the mass matrix {Eq.~\ref{Zmatrix}} one obtains $n+1$ (usually) massive
eigenstates $Z_{\alpha\mu}$ with mass $M_\alpha$,
\beq
Z_{\alpha\mu}=\sum_{\beta=1}^{n+1} U_{\alpha\beta} Z_{\beta\mu}^0,
\eeql{eigenstates}
where $U$ is an orthogonal mixing matrix.
It is straightforward to show that the mass-squared eigenvalues
are always nonnegative.
From {Eq.~\ref{lagrangian}} and {Eq.~\ref{eigenstates}} $Z_{\alpha\mu}$ couples to
$\sum_\beta g_\beta  U_{\alpha\beta} J^\mu_\beta$.

The most studied case is $n=1$. Writing $Q_i\equiv Q_{2i}$, the mass matrix is
\beq
\begin{split}
M^2_{Z-Z'} &= \left(
\begin{array}{cc}
2 g_1^2 \sum_i t_{3i}^2 |\langle \phi_i \rangle|^2 &
 2 g_1 g_2 \sum_i t_{3i}Q_i |\langle \phi_i \rangle|^2 \\
 2 g_1 g_2 \sum_i t_{3i}Q_i |\langle \phi_i \rangle|^2& 2 g_2^2 \sum_i Q_i^2 |\langle \phi_i \rangle|^2
\end{array}
\right)\\
& \equiv  \left(
\begin{array}{cc}
 M_{Z^0}^2 & \Delta^2 \\  \Delta^2 & M_{Z'}^2
\end{array} \right).
\end{split}
 \eeql{massmatrix}
As an example, many \uprm\ models involve  an $SU(2)$ singlet, $S$, and  two Higgs doublets, 
\beq \phi_u= \douba{\phi^0_u}{\phi^-_u}, \qquad \phi_d= \douba{\phi^+_d}{\phi^0_d},
\eeql{twohiggs}
 with \uprm\ charges $Q_{S,u,d}$. Then 
\beq \begin{split}
M_{Z^0}^2 &=  \frac{1}{4} g_1^2 (|\nu_u|^2+|\nu_d|^2)   \\
 \Delta^2 &=  \oh g_1 g_2 (Q_u |\nu_u|^2-Q_d |\nu_d|^2) \\
M_{Z'}^2 &= g_2^2( Q_u^2 |\nu_u|^2 + Q_d^2|\nu_d|^2+Q_S^2 |s|^2), 
\end{split} \eeql{examplepars}
\label{example}
where $\nu_{u,d} \equiv \sqrt{2} \langle  \phi^0_{u,d} \rangle $, $s=\sqrt{2}  \langle S \rangle $,
and $\nu^2=(|\nu_u|^2+|\nu_d|^2)\sim (246 {\rm\ GeV})^2$.

The eigenvalues of a general   $M^2_{Z-Z'}$ are
\beq
M^2_{1,2}=\oh \left[  M_{Z^0}^2+ M_{Z'}^2\mp \sqrt{(M_{Z^0}^2- M_{Z'}^2 )^2+ 4 \Delta^4} \right],
\eeql{eigenvalues}
and $U$ is the rotation
\beq
U=\left(
\begin{array}{cc}
  \cos \theta   & \sin \theta  \\
- \sin \theta  &     \cos \theta   
\end{array}
\right),
\eeql{rotation}
with
\beq
\theta= \oh \arctan \left( \frac{2\Delta^2}{M_{Z^0}^2- M_{Z'}^2} \right).
\eeql{eigenvectors}
$\theta$ is related to the masses by
\beq
\tan^2\theta =\frac{M_{Z^0}^2-M_1^2}{M_2^2-M_{Z^0}^2}.
\eeql{mixing1}

 An important limit is $M_{Z'} \gg (M_{Z^0}, |\Delta|)$, which typically occurs because an $SU(2)$ singlet field (such as $S$ in the example) has a large VEV and contributes only to $M_{Z'}$. One then has
\beq
M_1^2 \sim M_{Z^0}^2 - \frac{\Delta^4}{M_{Z'}^2}\ll M_2^2, \qquad M_2^2 \sim M_{Z'}^2
\eeql{eigenvalueslimit}
and
\beq
\theta \sim -\frac{\Delta^2}{M_{Z'}^2}\sim C \frac{g_2}{g_1} \frac{M_1^2}{M_2^2}
{\rm \ with \ } C=- \frac{ \sum_i t_{3i}Q_i |\langle \phi_i \rangle|^2}{\sum_i t_{3i}^2 |\langle \phi_i \rangle|^2 }.
\eeql{mixing2}
$C$ is model dependent, but typically $|C|  \lesssim {\cal O}(1)$. From {Eq.~\ref{mixing1}}-{\ref{mixing2}}
one sees that both $|\theta|$ and the downward shift $(M_{Z^0}-M_1)/M_{Z^0}$ are of
order $M_1^2/M_2^2$. Generalizations of these results for $n>1$ extra $U(1)$'s are given
in~\cite{Langacker:1984dp}.

\subsection{Anomalies and Exotics}\label{sec:anomalies}
A symmetry is chiral if it acts differently on the left and right handed fermions, and non-chiral (or vector)
otherwise. Thus, a chiral  \uprm\ has $Q_{\alpha f}\ne -Q_{\alpha f^c}$ for at least one $f$,
which is also referred to as chiral.
Even for a chiral symmetry, some of the fermions may be non-chiral.
If a given fermion pair is non-chiral with respect to all of the symmetries then an elementary mass term $-L_m=m_f \bar f_L f_R + h.c.$ is allowed,
where $m_f$ could be arbitrarily large. Such a term is forbidden for a chiral fermion, whose mass
is only generated when the symmetry is broken.  For example, if the symmetries allow the
Yukawa coupling
\beq
-L_{Yuk}=\lambda_f \varphi  \bar f_L f_R + h.c.,
\eeql{eqn:yukawa}
where  $\varphi$ is charged under the symmetry, then an effective mass  $\lambda_f\langle  \varphi \rangle $ is generated when $\varphi$ acquires a VEV. Assuming $\lambda_f \lesssim 1$,  $m_f$ cannot be larger than
the symmetry breaking scale $\langle  \varphi \rangle $.  In the SM, the ordinary quarks and
leptons are chiral under both $SU(2)$ and $U(1)_Y$, and $\varphi$ is the Higgs doublet. Similar
constraints apply to new fields occurring in \uprm\ models, which are frequently
{\em quasi-chiral}, i.e., non-chiral under the SM but chiral under \uprm.

Consistency of a  low-energy gauge theory requires the absence of triangle anomalies, including mixed
gauge-gravitational ones\footnote{See Section~\ref{sec:models} for the role of anomalous \uprm~s.}.
For the SM, the non-trivial conditions are
\beq
\begin{split}
\sum_f Y_f&=0, \qquad \sum_f Y_f^3=0,\\ \sum_{f\in 3,3^*} Y_f&=0, \qquad \sum_{f\in2} Y_f=0,
\end{split}
\eeql{smanomaly}
where the sum extends over all left-handed fermions ($Y_f=y_{f_L}$) {\em and} antifermions ($Y_{f^c}=-y_{f_R}$). The first condition
is the mixed anomaly; the sum is over color triplets and antitriplets in the third  $[SU(3)^2Y]$
condition; and the sum is over $SU(2)$ doublets in the last $[SU(2)^2Y]$ condition. The sum includes
counting factors of 3 for families, 3 for color triplets, and 2 for $SU(2)$ doublets, since $SU(3)$ and $SU(2)$
commute with hypercharge and additional \uprm~s. For example, the second $[Y^3]$ condition is
\beq
3[6Y_Q^3+2Y_L^3+3Y_{u^c}^3+3Y_{d^c}^3+Y_{e^c}^3]=0,
\eeql{y3}
where $Q$ and $L$ refer respectively to quark and lepton doublets.
This is satisfied by a cancellation between quark and lepton terms.
One also requires the absence of an $[SU(3)^3]$ anomaly. In the SM this is achieved automatically
because there equal numbers of quarks and antiquarks.
With an additional \uprm\ with charge $Q_2$, there are additional
conditions obtained from {Eq.~\ref{smanomaly}} by replacing $Y$ by $Q_2$. There are also mixed
$[Y \uprm]$ conditions $\sum_f Y Q_2^2=\sum_fY^2Q_2=0$. For $n>1$ additional \uprm\ there 
are similar conditions for every $Q_\beta, \beta \ge 2$ as well as $\sum_f Y Q_\alpha Q_\beta=\sum_f
Q_\alpha Q_\beta Q_\gamma=0$. All of these sums include any extra chiral fermions in the theory, such as
the superpartners of Higgs scalars in supersymmetry. Non-chiral fermion pairs cancel. 

Even for a single \uprm\ it is easy to show that the anomaly conditions cannot be satisfied by
the SM fermions alone if the \uprm\ charges are the same for all three families, except for the
trivial case $Q_2=0$.
( $Q_2=cY$ is also possible, but this is equivalent to $Q_2=0$ after performing a
rotation on $B_\mu$ and $Z^0_{2\mu}$.)
Thus, almost all \uprm\ constructions involve
additional fermions, known as exotics. These may be singlets under the SM gauge group, such as a singlet right-handed neutrino,
or they may carry nontrivial SM quantum numbers. Precision electroweak constraints strongly restrict,
but don't entirely exclude, the possibilities of new fermions chiral under \stoy~\cite{Yao:2006px},
so such exotics are usually assumed to be quasi-chiral, e.g.,
both left and right-handed components might be $SU(2)$ doublets, or both might be singlets.
A typical example is a new $SU(2)$-singlet heavy down-type quark $D$ with $q_D=-1/3$
and its partner $D^c$.

One can introduce vector pairs that are charged but non-chiral under both the SM and \uprm.
These do not contribute to the anomaly conditions, but contribute to the renormalization group
equations (RGE) for the gauge couplings, and may also be relevant to the decays of exotics.

If two \uprm\ charges $Q_{\alpha,\beta}$ (one of these can be $Y$) are both generators of a
simple underlying group, then one expects them to be orthogonal, 
i.e.,  $\sum_{f }  Q_{\alpha f} Q_{\beta f}=0$ for $\alpha\ne\beta$, with a corresponding condition
for the scalar charges. However, this condition need not hold without such an embedding or
for a more complicated one, or it could be violated due to kinetic mixing (Section~\ref{sec:kinetic}). Furthermore, all fermions, including the non-chiral ones, contribute to the orthogonality condition. 
In particular, an apparent violation of orthogonality could be due to the fact that the
contributions of a very heavy vector pair (or of heavy scalars) have not been taken into account.
There is always some freedom to perform rotations on the gauge fields $Z^0_{\alpha\mu}$,
e.g., to make  the \uprm\ charges orthogonal (at least with respect to the fermions, in a
nonsupersymmetric theory). However, such  a rotated basis may not coincide with either
the mass or kinetic eigenstates.

\subsection{Kinetic Mixing}\label{sec:kinetic}
The  most general  kinetic energy term for the two gauge bosons $Z^0_{\alpha\mu}$ and
$Z^0_{\beta\mu}$ in $U(1)_\alpha \x U(1)_\beta$ is
\beq
L_{kin}\ra-\frac{c_\alpha}{4} F^{0\mu\nu}_\alpha F^0_{\alpha\mu\nu}-\frac{c_\beta}{4} F^{0\mu\nu}_\beta F^0_{\beta\mu\nu}
-\frac {c_{\alpha\beta}}{2} F^{0\mu\nu}_\alpha F^0_{\beta\mu\nu},
\eeql{eqn:kinetic}
where $F^0_{\alpha\mu\nu}=\partial_\mu Z^0_{\alpha\nu}   - \partial_\nu Z^0_{\alpha\mu}$.
One can put the first two terms into canonical form $c_\alpha=c_\beta=1$ by rescaling the
fields, and take $c_{\alpha\beta}=\sin \chi$. Since $U(1)$  field strengths are invariant,
the  cross (kinetic mixing) term does not spoil the
gauge invariance~\cite{Holdom:1985ag}. Even if $\chi=0$ at tree level, it can be
generated by loop effects if there are particles in the theory that are simultaneously
charged under both $U(1)$'s~\cite{Holdom:1985ag,Matsuoka:1986ig,delAguila:1988jz,Foot:1991kb,delAguila:1995rb,Babu:1996vt,Babu:1997st}.
The mixing term can be cast as a cross term in the renormalization group
equations (RGE) for the gauge couplings, with a coefficient proportional to
$\sum_{m_f < \mu} Q_{\alpha f} Q_{\beta f}$, where $\mu$ is the RGE scale, with corresponding contributions from scalars. Even for orthogonal charges
the sum at lower mass scales may be nonzero due to the decoupling of heavy
particles. Such RGE effects are usually of order a few \% in $\chi$,
but could be larger if there are many decoupled 
states~\cite{Babu:1996vt}. A non-zero $\chi$ can also be generated by
string loop effects in superstring theory. These contributions are
small in the heterotic constructions considered in~\cite{Dienes:1996zr}. 
However, if one of the $U(1)$ factors is broken in a hidden sector at a
large scale, the associated $D$ terms could propagate this scale to
the ordinary sector by kinetic mixing, destabilizing the supersymmetry breaking scale
and leading to negative mass-square scalars~\cite{Dienes:1996zr}. 

Now, consider the consequences of kinetic mixing for a single extra \uprm,
i.e., $\alpha=1, \beta=2$. $L_{kin}$ can be put in canonical form (for $c_{1,2}=1,
c_{12}=\sin\chi$) by defining 
\beq
\douba{Z^0_{1\mu}}{Z^0_{2\mu}}= 
\left(
\begin{array}{cc} 1 &   -\tan \chi   \\ 0 &   1/\cos \chi\end{array}\right)
\douba{\hat Z^0_{1\mu}}{\hat Z^0_{2\mu}}
\equiv V \douba{\hat Z^0_{1\mu}}{\hat Z^0_{2\mu}},
\eeql{eqn:can}
where $V$ is non-unitary. In the new $\hat Z$ basis, the mass matrix  in {Eq.~\ref{massmatrix}}
becomes $V^T M^2_{Z-Z'} V$, which can be diagonalized by an orthogonal matrix $U^T$. Similarly, the interaction term in {Eq.~\ref{lagrangian}} becomes
\beq
\begin{split}
\left(g_1 J^\mu_1\quad       g_2 J^\mu_2\right)
\douba{ Z^0_{1\mu}}{Z^0_{2\mu}}& \equiv {\cal J}^T \douba{ Z^0_{1\mu}}{Z^0_{2\mu}}
\ra {\cal J}^T V \douba{\hat Z^0_{1\mu}}{\hat Z^0_{2\mu}}\\
&={\cal J}^T V U^T \douba{ Z_{1\mu}}{Z_{2\mu}},
\end{split}
 \eeql{eqn:int}
 where $Z_{1,2}$ are the mass eigenstates. These transformations are analyzed in detail, in, e.g.,~\cite{Babu:1997st}.
 The essential feature can be seen for $\Delta^2=0$ in {Eq.~\ref{massmatrix}}, for which
\begin{multline} \label{eqn:kinmasses}
V^T M^2_{Z-Z'} V =\\
\left( \begin{array}{cc}  M_{Z^0}^2 &   - M_{Z^0}^2\tan \chi   \\ - M_{Z^0}^2\tan \chi  &   
 M_{Z^0}^2\tan^2 \chi + M_{Z'}^2/\cos^2 \chi\end{array}\right).
\end{multline}
One sees immediately that  for $M_{Z^0}^2=0 $ there is a zero eigenvalue, even for large $\chi$,
i.e.,  any shift in the lighter mass induced by  kinetic mixing is proportional to the
light mass and therefore small. In fact, for $|M_{Z^0}^2| \ll |M_{Z'}^2|$, $\Delta^2=0$,
and $|\chi| \ll 1$ one has $M_1^2\sim M_{Z^0}^2-M_{Z^0}^4 \chi^2/M_{Z'}^2$, a negligible
shift. The only significant effect in this limit is that the couplings become
\beq
 g_1 J^\mu_1 Z_{1 \mu}+  (g_2 J^\mu_2 -g_1 \chi J^\mu_1) Z_{2 \mu},
\eeql{eqn:shift}
i.e., the coupling of the heavy boson is shifted to include a small component proportional
to $J_1$. The light boson couplings are not affected to this order. One must
still include the further effects of mass mixing ($\Delta^2\ne 0$). 
For $|\Delta^2| \lesssim |M_{Z^0}^2| \ll |M_{Z'}^2|$ and $|\chi| \ll 1$,
\beq
M_1^2\sim M_{Z^0}^2-\frac{(\Delta^2-M_{Z^0}^2 \chi)^2}{M_{Z'}^2}
\sim M_{Z^0}^2-\hat \theta^2 M_{Z'}^2,
\eeql{kmzmass}
where 
\beq
 \hat \theta \equiv \frac{(-\Delta^2+M_{Z^0}^2 \chi)}{M_{Z'}^2}.
\eeql{kmzmix}
This is of the same form as {Eq.~\ref{eigenvalueslimit}} except that the effective mixing angle
is shifted from {Eq.~\ref{mixing2}} by the kinetic mixing. The interactions are just the rotation by
$\hat \theta$ of those in {Eq.~\ref{eqn:shift}}.
 
In a supersymmetric theory the charges in the $U(1)_2$ $D$ terms
are also shifted, $g_2Q_2\ra g_2Q_2-g_1 \chi Q_1$.  There can also be kinetic mixing 
between the $U(1)'$ gauginos~\cite{Suematsu:1998wm}, with consequences analogous to
those for the gauge bosons.

\subsection{One and Two Higgs Doublets, Supersymmetry,  and the $\mu$ Problem}\label{twodoublet}
\subsubsection{Higgs Doublets}\label{higgsdoublets}
The standard model involves a single Higgs doublet  $\phi= \douba{\phi^+}{\phi^0}$,
which has Yukawa couplings (ignoring family indices)
\beq\begin{split}
-L_{Yuk}&=h_d\bar Q_L \phi d_R + h_u \bar Q_L \widetilde \phi u_R \\
&+ h_e\bar L_L \phi e^-_R + h_\nu \bar L_L \widetilde \phi \nu_R+h.c., 
\end{split}
\eeql{lyuk}
where 
$Q_L\equiv \douba{u_L}{d_L}$,  $L_L\equiv \douba{\nu_L}{e^-_L}$, and $\nu_R$ is the 
right-handed (SM-singlet) neutrino. The tilde field is defined by
\beq
\widetilde \phi\equiv i \sigma^2 \phi^\ast= \douba{\phi^{0\ast}}{-\phi^-}, \eeql{tildephi}
where $\sigma^2$ is the second Pauli matrix.
It is essentially the Hermitian conjugate of $\phi$, but transforms as a $\bf 2$ rather than a $\bf 2^*$
under $SU(2)$, and has $y=-1/2$. A single doublet suffices for the SM, but in many extensions,
including supersymmetry and many \uprm\ models, the $\widetilde \phi$ couplings are not allowed. One must introduce
a second (independent) doublet  $\phi_u$, as in {Eq.~\ref{twohiggs}}, which plays the role of 
$\widetilde \phi$, while $\phi_d$ plays that of $\phi$.

In supersymmetric models it is convenient to work entirely in terms of (left) chiral superfields,
such as $Q, L, u^c, d^c, e^+$, and the SM singlet $\nu^c$ which is conjugate to $\nu_R$ (we do not distinguish between chiral superfields and their
components in our notation -- the context should always make the  meaning clear).
Furthermore, supersymmetry (and anomaly constraints) require two Higgs doublets
 $H_u=\douba{H_u^+}{H_u^0}$ and $H_d=\douba{H_d^0}{H_d^-}$
with $Y_{H_{u,d}}=\pm1/2$, defined so that the MSSM superpotential is 
\beq
\begin{split}
W&=\mu H_u H_d
-h_d Q H_d d^c + h_u Q H_u u^c\\
&-h_e L H_d e^+ + h_\nu L H_u \nu^c.
\end{split}
 \eeql{wmssm}
Doublets are contracted according to  $H_u H_d \equiv \epsilon^{ab} H_{ua} H_{db}$, etc., where 
$\epsilon^{12}=-\epsilon^{21}=1$. The two sets of Higgs doublets  are related by $H_{u,d}=\mp \widetilde \phi_{u,d}$. The superpotential in {Eq.~\ref{wmssm}}  assumes that $R$-parity is conserved. Some \uprm\ models enforce this automatically, as will be mentioned in {Section~\ref{sec:anomfree}}.

\subsubsection{Non-Holomorphic Terms}\label{nonholomorphic}
In some \uprm\ extensions of the MSSM, some of the Yukawa couplings in {Eq.~\ref{wmssm}}
may be forbidden by the \uprm\ gauge symmetry. In some cases, however, the
operators involving the {\em wrong} Higgs field, such as  $Q \widetilde H_u d^c$ or $L \widetilde H_u e^+$, may be \uprm\ invariant. Such non-holomorphic operators are not allowed in $W$ by supersymmetry,
 but could be present in the  K\"ahler potential, where they would lead to corresponding
 non-holomorphic soft terms~\cite{Borzumati:1999sp}  for the scalar squarks and sleptons.
These then lead to fermion masses at one loop by gluino or neutralino exchange.
However, in most supersymmetry breaking schemes it is difficult to generate a large enough
effective Yukawa~\cite{Martin:1997ns}, because the
non-holomorphic soft terms have an additional suppression 
(compared to the usual soft SUSY breaking scale of  $M_{SUSY}\sim 1$ TeV)
of $M_{SUSY}/M_{med}$,
where  $M_{med}\gg M_{SUSY}$
is the SUSY mediation scale (such as the Planck scale for supergravity mediation).

\subsubsection{The $\mu$ Problem}\label{muproblem}
One difficulty with the MSSM is the $\mu$ problem~\cite{Kim:1983dt}, i.e., the 
supersymmetric Higgs mass $\mu$ in {Eq.~\ref{wmssm}} could be arbitrarily large,
but phenomenologically needs to be of the same order as the soft supersymmetry breaking terms.
In many supersymmetric  \uprm\ models this problem is solved because an elementary $\mu$ term
is forbidden by the \uprm,
but a trilinear $W_\mu= \lambda_S S H_u H_d$ is allowed, where $S$ is a singlet under the SM but
charged under the \uprm. Then,  a dynamical effective
$\mu_{eff}= \lambda_S \langle S \rangle$ is generated that is related to the scale of \uprm\ 
breaking~\cite{Cvetic:1995rj,Suematsu:1994qm,Cvetic:1997ky}, as will be further discussed in Section~\ref{sec:mu}. This mechanism
can also be associated with discrete or other symmetries~\cite{Accomando:2006ga}. An alternative
solution, the Giudice-Masiero mechanism, generates $\mu$ through a nonrenormalizable operator in the K\"ahler potential~\cite{Giudice:1988yz}.
It is especially useful when an elementary $\mu$ term is allowed by the low energy symmetries of the theory, but
is forbidden by the underlying string construction. This mechanism can also be used to generate
mass for vector pairs in \uprm\  theories.

\section{Models}\label{sec:models}
There are enormous numbers of \uprm\ models, and it is only possible to touch on the
major classes and issues here. The models are distinguished by: (a)
the coupling constants $g_\alpha$, which are often assumed to be of electroweak
strength, but could be larger or smaller. (b) The \uprm\ breaking scale. In some scenarios
this is arbitrary, with no good reason to expect it to be around the TeV scale. However, in
supersymmetric models it is usually at the TeV scale, unless the breaking is associated with
an $F$ and $D$ flat direction, when it could be much larger. The TeV scale is also expected 
when the \uprm\ is associated with alternative models of electroweak breaking.
String constructions usually imply some \zpr~s close to the string scale, and often involve
lighter ones as well. Finally, a \zpr\ could actually be lighter than the electroweak scale
if its couplings to the SM fields are small. (c) Other critical issues are the charges of the
SM fermions and Higgs doublet, and whether the fermion charges are family universal; the type of scalar responsible for the \uprm\ breaking; whether
additional exotic fields are needed to cancel anomalies; whether the theory is supersymmetric
(so that the Higgs superpartners must be included in the anomaly considerations); whether
the Yukawa couplings of the ordinary fermions are allowed by the \uprm\ symmetry; and whether
other couplings, such as those associated with the supersymmetric $\mu$ parameter, $R$-parity
violation, and Majorana neutrino masses are allowed.

\subsection{Canonical Examples}\label{sec:canonical}
\subsubsection{The sequential model}\label{sec:sequential}
The {\em sequential\/} $Z_{SM}$ boson is defined to have the same 
        couplings to fermions as the SM $Z$ boson. The $Z_{SM}$ is not expected
        in the context of gauge theories unless it has different couplings to 
        exotic fermions, or if it occurs as an excited state of the ordinary $Z$
        in models with extra dimensions at the weak scale. However, it serves as a useful
        reference case when comparing constraints from various sources.
\subsubsection{Models based on $T_{3R}$ and $B - L$}\label{lrmodels}

One of the simplest and most common classes of models involves
$SU(2) \x U(1)_{3R} \x U(1)_{BL}$, where the $U(1)_{3R}$ generator $T_{3R}$ is $\oh$ for $(u_R, \nu_R)$, $-\oh$
for $(d_R, e^-_R)$, and $0$ for $f_L$; and the $U(1)_{BL}$ generator is
$T_{BL}\equiv \oh (B-L)$, where $B(L)$ is baryon (lepton) number; and
$\nu_R$ are right-handed neutrinos. (See Table~\ref{tab:t3rtbl}.) $T_{3R}$ and $T_{BL}$ are
related to weak hypercharge by $Y=T_{3R}+T_{BL}$. 
$T_{3R}$ occurs in left-right symmetric models based on the
group $G_{LR}\equiv SU(2)_L \x SU(2)_R \x U(1)_{BL}$ (for a review, see~\cite{Mohapatra:737303})
and in $SO(10)$ models (which contain $G_{LR}$)~\cite{Langacker:1980js,Hewett:1988xc}.
The Higgs doublet $\phi$ can be assigned  $T_{3R}=\oh$ and $T_{BL}=0$. However,
in the $G_{LR}$ or $SO(10)$ embeddings (or in supersymmetric versions), there are
two Higgs doublets, $\phi_{u,d}$, as defined in {Eq.~\ref{twohiggs}},
with $T_{3R}=\mp \oh$. All of these versions are anomaly-free after including the three $\nu_R$.

For these models, the fermion neutral current couplings are
\beq
-L_{\rm NC}=gJ_{3L} W_{3L} + g_R J_{3R} W_{3R}
+ g_{BL} J_{BL} W_{BL},
\eeql{t3rbl}
where $J_{3L}\equiv J_3$, $J_{3R}$ and $J_{BL}$ are the currents corresponding to 
$T_{3R}$ and $T_{BL}$,  the  $g$'s and $W$'s are the coupling constants and
gauge bosons, and the Lorentz indices have been suppressed.

We anticipate that $U(1)_{3R} \x U(1)_{BL}$ will be broken to $U(1)_Y$ at a
scale $M_{\zpr}\gg M_{Z^0},$ so it is convenient to first transform the gauge bosons
$W_{3R}$ and $W_{BL}$ to a new basis $B$ and $Z^0_2$, where $B$ is identified with
the SM $U(1)_Y$ boson,
\beq
\begin{split}
-L_{\rm NC}&=gJ_{3L} W_{3L} + g' J_Y B
+ g_2 J_2 Z^0_2 \\ &= eJ_{em} A+ g_1 J_1 Z^0_1 + g_2 J_2 Z^0_2,
\end{split}
\eeql{t3rblrot}
as in {Eq.~\ref{lagrangian}}. 
Let us first assume that the gauge kinetic terms are canonical, i.e., with unit strength and no 
kinetic mixing, so that  orthogonal transformations on the three gauge bosons will leave the kinetic
terms invariant. Taking $B\equiv \cos \gamma\, W_{3R}+\sin \gamma\, W_{BL}$ and
choosing $\gamma$ so that $B$ couples to $g' Y$, 
one finds $1/g'^2=1/g_R^2+1/g_{BL}^2$, and that the
orthogonal gauge boson $ Z^0_2= \sin \gamma\, W_{3R}-\cos \gamma\, W_{BL}$ couples to the current
$J_2$ associated with the charge
\beq Q^{LR} =\sqrt{\frac{3}{5}} \left[  \alpha T_{3R}-\frac{1}{\alpha} T_{BL} \right],
\eeql{J2LR}
where 
\beq \alpha=\frac{g_R}{g_{BL}}=\sqrt{\kappa^2 \cot^2 \theta_W-1}, 
\eeql{alphadef}
with $\kappa \equiv g_R/g$.
The coupling has been normalized to 
\beq g_2=\sqrt{\frac{5}{3}} g \tan \theta_W\sim 0.46\eeql{standardcoupling}
for later convenience.

One interesting case is when $G_{LR}$ survives down to the TeV scale. 
This is usually studied assuming a left-right symmetry under the interchange of the
two $SU(2)$ factors~\cite{Mohapatra:737303}, in which case $g_R=g$
and $\alpha\sim 1.53$ for $\sinn\sim 0.23$. Two forms of the model are
often considered. In both cases, the Higgs doublets $\phi_{u,d}$ responsible for fermion
mass transform as $(2,2)_0$ under $SU(2)_L\x SU(2)_R$, where the subscript is the 
$T_{BL}$ charge. In one class, an additional
doublet pair $\delta_{R,L}$  transforming as $(1,2)_{1/2}+(2,1)_{1/2}$ is introduced, with the
VEV of $\delta^0_R$ breaking $G_{LR}$ to the SM. In the other, one instead introduces
a triplet pair $\Delta_{R,L}$  transforming as $(1,3)_{1}+(3,1)_{1}$. The $\Delta^0_R$
VEV not only breaks $G_{LR}$ but also leads to a large Majorana mass for  the
$\nu_R$ and therefore a small $\nu_L$ mass by the seesaw mechanism~\cite{Mohapatra:1979ia}.
The low-scale left-right model also implies a new $W^\pm_R$ which couples to right-handed
currents and can mix with the SM $W^\pm$. Strategies for determining the symmetry breaking pattern
were described in~\cite{Cvetic:1991kh}, and limits on the charged sector masses and mixings for general models without left-right symmetry are given in~\cite{Langacker:1989xa,Yao:2006px}.

The simple forms of the (supersymmetric) left-right model are not consistent with gauge
unification unless the $SU(2)_R$ breaking occurs at a much higher scale (e.g.,
$10^{12}$ GeV).
Such a large scale is also required by current allowed ranges for the neutrino
masses in the triplet versions. In some cases, the
initial breaking can leave  $U(1)_{3R} \x U(1)_{BL}$  unbroken.
Realistic  $SO(10)$ breaking patterns suggest $\alpha$
in the range $0.7-0.9$~\cite{Robinett:1981yz}. An important special case is the
$\chi$ model, which occurs when $SO(10)$ breaks directly to $SU(5)\x U(1)_\chi$.
This corresponds to {Eq.~\ref{J2LR}} for $\kappa=1$ and
$\sinn=3/8$ (which is the value predicted by $SU(5)$ {\em at the unification scale}), leading to $\alpha=\sqrt{2/3}\sim 0.82$.

A generalization of this type of model is based on $SU(2) \x U(1)_{Y} \x U(1)_{2}$,
where $Q_2$ is a linear combination 
\beq
Q^{YBL}=a Y + b T_{BL}\equiv b(zY+T_{BL}), 
\eeql{ybldef}
where $b\ne 0$.
It is convenient to normalize $b$ so that the coupling $g_2$ 
is given by {Eq.~\ref{standardcoupling}}, or alternatively one can choose $b=1$ and take $g_2$
to be arbitrary. The $U(1)_{3R} \x U(1)_{BL}$ limit in {Eq.~\ref{J2LR}} corresponds to choosing
$b^2 z (1+z)=-3/5$ with $\alpha \equiv \sqrt{5/3} bz$.
$Q^{YBL}$ is anomaly free for the standard model fermions (including $\nu_R$)~\cite{Weinberg:283223}. 
$Y$ and $ Q^{YBL}$ are non-orthogonal (i.e.,  $\sum_f Y_f Q^{YBL}_{f}\ne 0$ when summed over a family of the known left-handed
fermions and antifermions), except for the special case of  $U(1)_{3R} \x U(1)_{BL}$,
but it could come about
by a more general embedding of the generators or by (possibly large) kinetic mixing, as discussed in Section~\ref{sec:anomalies}.
 The pure $B-L$
model ($z=0$) is often studied phenomenologically, and has the property that the ordinary
Higgs doublets do not induce $Z-Z'$ mixing. The models in this class have been
systematically discussed in~\cite{Appelquist:2002mw}, including generalizations
with an arbitrary number of $\nu_R$ with nonuniversal charges.

This entire class of models based on $T_{3R}$ and $T_{BL}$ (or $Y$ and $T_{BL}$)
is perhaps less interesting in a supersymmetric context, because the two supersymmetric
Higgs doublets $H_{u,d}$ form a vector pair with $T_{3R}=\pm \oh$ and $T_{BL}=0$.
Therefore,  an elementary $\mu$ term in {Eq.~\ref{wmssm}}  is {\em not} forbidden by the extra \uprm.
Similar difficulties apply to the SM singlet supermultiplets that are needed to break the \uprm, since
they would most likely be introduced as non-chiral vector pairs to avoid anomalies. (One could instead
give large VEVs  to the scalar partners of the $\nu^c$, but this would break $R$-parity and would be
challenging for neutrino phenomenology.)
\begin{table}[ht]
\caption{Charges of the left-chiral components of the fermions in the models
based on  $T_{3R}$ and $T_{BL}=(B - L)/2$. The charges are normalized so that 
 $ g_2=\sqrt{\frac{5}{3}} g \tan \theta_W$. 
  $Q^{LR}$ is defined in {Eq.~\ref{J2LR}}, and 
$Q^{YBL}\equiv b(zY+T_{BL}) $. 
  $Q^{LR}$
is a special case of $Q^{YBL}$ for $b^2 z (1+z)=-3/5$.
$\alpha$ and $(b,z)$ are free parameters, with $\alpha=1.53$ for left-right
symmetry and $\alpha \sim 0.7-0.9$ for most $SO(10)$ models.
\label{tab:t3rtbl}}
\begin{center}
\begin{tabular}{|c| c| c| c| c| c|}
\hline
  & $T_{3R}$ & $T_{BL}$ & $Y$ & $\sqrt{\frac{5}{3}}Q^{LR}$ & $\frac{1}{b} Q^{YBL}$ \\
  \hline
$Q$ & $0$  & $\frac{1}{6}$& $\frac{1}{6}$ & $-\frac{1}{6\alpha}$&$\frac{1}{6}(z+1)$\\
$u^c_L$ & $-\frac{1}{2}$  & $-\frac{1}{6}$ & $-\frac{2}{3}$ & $-\frac{\alpha}{2}+\frac{1}{6\alpha}$&$-\frac{2}{3}z- \frac{1}{6}$\\
$d^c_L$ & $\frac{1}{2}$  & $-\frac{1}{6}$& $\frac{1}{3}$ & $\frac{\alpha}{2}+\frac{1}{6\alpha}$&$\frac{1}{3}z- \frac{1}{6}$\\
$L_L$ & $0$  & $-\frac{1}{2}$ & $-\frac{1}{2}$& $\frac{1}{2\alpha}$&$-\frac{1}{2}(z+1)$\\
$e^+_L$ & $\frac{1}{2}$  & $\frac{1}{2}$ & $1$& $\frac{\alpha}{2}-\frac{1}{2\alpha}$& $z+\frac{1}{2}$\\\
$\nu^c_L$ & $-\frac{1}{2}$  & $\frac{1}{2}$& $0$& $-\frac{\alpha}{2}-\frac{1}{2\alpha}$& $\frac{1}{2}$\ \\
\hline
\end{tabular}
\end{center}
\end{table}

\subsubsection{The $E_6$ models}\label{sec:e6model}
Many \zpr\ studies focus on the two extra \uprm~s which occur in
the decomposition of the $E_6$ GUT~\cite{Robinett:1982tq,Langacker:1984dc,Hewett:1988xc},
i.e., $E_6 \ra SO(10)\x U(1)_\psi$ and $SO(10)\ra SU(5) \x U(1)_\chi$. We consider them only as
simple examples of anomaly-free \uprm\ charges and exotic fields, and do not assume a full
underlying grand unified theory.
In $E_6$, each family of left-handed fermions is promoted to a fundamental $\bf 27$-plet, 
which decomposes under
$E_6\ra SO(10)\ra SU(5)$ as 
\beq
{\bf 27}\ra {\bf 16}+{\bf 10}+{\bf 1}\ra ({\bf 10}+{\bf 5^\ast}+{\bf 1})+({\bf 5}+{\bf 5^\ast})+{\bf 1},
\eeql{e6decomp}
as shown in Table~\ref{tab:e6decomp}.
In addition to the standard model fermions, each $\bf 27$-plet contains two standard model singlets,
$\nu^c$ and $S$ (which may be charged under the \uprm). The $\nu^c$ may be interpreted as the conjugate of the right-handed neutrino. There is also an exotic
color-triplet quark $D$ with charge $-1/3$ and its  conjugate $D^c$,
both of which are $SU(2)$ singlets,
and a pair of color-singlet $SU(2)$-doublet exotics,
$H_u=\douba{H_u^+}{H_u^0}$ and $H_d=\douba{H_d^0}{H_d^-}$
with $y_{H_{u,d}}=\pm1/2$. $H_d$ transforms the same way as $H^c_u\equiv \widetilde H_u$, the
(tilde) conjugate of $H_u$ under the SM. The exotic fields are all therefore singlets or non-chiral under the standard model, 
but may be  chiral under the \uprm.

The $E_6$ models can be considered in both non-supersymmetric and supersymmetric versions.
In the supersymmetric case, the scalar partners of the $S$ and $\nu^c$ can develop VEVs to
break the \uprm\ symmetry, though the latter (as well as a VEV for the scalar partner of the $\nu$)
would break $R$-parity and may be problematic for neutrino phenomenology.
Similarly, the scalar partners  of one $H_{u,d}$ pair can be interpreted as the
two Higgs doublets of the MSSM.
The two additional $H_{u,d}$ families
may be interpreted either as additional Higgs pairs or as exotic-leptons
($H_d$ has the same SM quantum numbers as an ordinary lepton doublet, while $H_u$
would be conjugate to a right-handed exotic doublet).

 Table~\ref{tab:e6decomp}
also lists the $U(1)_{\chi}$ and $U(1)_{\psi}$ charges of the
$\bf 27$-plet. 
 By construction,
the fields in an irreducible representation of $SO(10)$ ($SU(5)$) all carry
the same $\psi$ ($\chi$) charges.
Most studies assume that only one \zpr, coupling to the linear combination 
\beq Q(\theta_{E_6})=\cos\theta_{E_6}
Q_\chi+\sin\theta_{E_6} Q_\psi, \eeql{qthetae6}
where $0\le\theta_{E_6}<\pi$ is a mixing angle, is relevant at low energies. 
(One can also include a kinetic mixing correction $-\epsilon Y$ to the effective charge, as
in {Eq.~\ref{eqn:shift}}).
As discussed in Section~\ref{lrmodels}, the $\chi$ model ($\theta_{E_6}=0$) is a special case of the  $T_{3R}$ and $B - L$
models, supplemented with additional exotic fields in the $\bf 10 +1$ of $SO(10)$.
Since the latter are non-chiral in this case they may be omitted, or one or more $\bf 10$'s
may be introduced as Higgs fields. The $\psi$ model ($\theta_{E_6}=\pi/2$), on the other hand,
has chiral exotics and requires the three full $\bf 27$-plets. Using {Eq.~\ref{charges}} one sees that
the currents of the fields  in the $\bf 16$ and $\bf 10$ have purely axial 
couplings to the $Z_\psi$ (this only  holds for the  $\nu$ if it pairs with the $\nu^c$
to form a Dirac fermion). Another commonly studied case is the $\eta$ model,
$Q_{\eta}=\sqrt{\frac{3}{8}}Q_{\chi}-\sqrt{\frac{5}{8}}Q_{\psi}=-Q
(\theta_{E_6}=\pi-\arctan\sqrt{5/3}\sim 0.71\pi$), which occurs in Calabi-Yau compactifications of the heterotic string if $E_{6}$ breaks directly to a rank 5 group \cite{Witten:1985xc} via the Wilson line (Hosotani) mechanism. 
The inert model, $Q_I=-Q
(\theta_{E_6}=\arctan\sqrt{3/5}\sim 0.21\pi$), has a charge orthogonal to $Q_\eta$ and follows from an alternative $E_6$ breaking
pattern~\cite{Robinett:1982tq}.
In the neutral $N$ model  ($\theta_{E_6}=\arctan \sqrt{15}\sim 0.42\pi$)~\cite{Ma:1995xk,Barger:2003zh,Kang:2004ix,King:2005jy},  the $\nu^c$ has zero charge, allowing a large Majorana mass or
avoiding big bang nucleosynthesis constraints for a Dirac $\nu$, as discussed in Section~\ref{sec:neutrino}.
It essentially interchanges the assignments of the $S$ and $\nu^c$ and of the two
$\bf 5^\ast$ representations (which have the same standard model quantum numbers)
with respect to the $\chi$ model, and is basically the same as the
alternative left-right model in~\cite{Ma:1986we,Babu:1987kp}. 
The secluded sector model ($\theta_{E_6}=\arctan (\sqrt{15}/9)\sim 0.13\pi$)~\cite{Erler:2002pr} will be discussed in Section~\ref{sec:intermediate}.

The $E_6$ models allow the Yukawa couplings needed to generate masses for
the standard model and exotic fermions. In particular, in the supersymmetric case
the superpotential terms
\beq
\begin{split}
 W&=-h_d Q H_d d^c + h_u Q H_u u^c-h_e L H_d e^+ + h_\nu L H_u \nu^c\\
 &+ \lambda_S S H_u H_d + \lambda_D S D D^c
\end{split}
\eeql{super}
are all allowed, where family indices have been neglected. (In the non-SUSY case, two Higgs doublets, analogous to $H_u$ and $H_d$, are required.) 
From {Eq.~\ref{super}} we see that the $E_6$ models all allow a dynamical $\mu_{eff}$,
while an elementary $\mu$ is forbidden in all but the $\chi$ model.

The supersymmetric $E_6$ model with three $\bf 27$-plets can incorporate one or more pairs
of Higgs doublets $H_{u,d}$ in the $\bf 5+ 5^\ast$ pairs. However, that version of the model is not
consistent with the simple form of gauge unification observed in the MSSM for the
SM subgroup. That is because the complete extra $\bf 5+ 5^\ast$ multiplets give equal
contributions to the $SU(3)$, $SU(2)$, and $U(1)_Y$ $\beta$ functions at one loop,
so the unification conditions are similar to the MSSM with 3 families but no Higgs pair.
Unification can be restored by introducing an $H_u$ and $H_u^c$ pair from an
incomplete $\bf 27+27^\ast$ representation~\cite{Langacker:1998tc}. (The physical $H_u$ could either be
this one or from the complete $\bf 27$-plets.) This pair is completely non-chiral, so it does not introduce
any anomalies, but at the cost of introducing a rather arbitrary aspect to the model.
Also, there is no obvious reason (except perhaps the mechanism in~\cite{Giudice:1988yz}) why this extra pair should be at the electroweak or TeV
scale, reintroducing a form of the $\mu$ problem. Nevertheless, the unification of the SM
gauge couplings and the unification scale $M_X$ are then the same as in the
MSSM at one loop, though the value of the gauge coupling at $M_X$ is  increased because
of the extra exotics. 

If the \uprm\ really derives from an $E_6$-type GUT which breaks directly to $\sthtoy\x U(1)'$,
 one expects that $g_2=\sqrt{\frac{5}{3}}g'$ at the unification scale, where $\sqrt{\frac{5}{3}}g'$
is the GUT-normalized hypercharge coupling. Running down to the TeV scale,
this implies
\beq
g_2=\sqrt{\frac{5}{3}} g \tan \theta_W \lambda_g^{1/2},
\eeql{e6gauge}
where $ \lambda_g^{1/2}\sim 1$ up to a ($\theta_{E_6}$-dependent)
correction of a few \% due to the \uprm\ charge of the incomplete $\bf 27+27^\ast$.
{Eq.~\ref{e6gauge}} can be taken as a definition of $\lambda_g$ for an arbitrary model.
It is typically of order unity even for more complicated $E_6$ breaking patterns~\cite{Robinett:1982tq},
and was taken to be unity by construction for the $G_{LR}$ model.

In a full $E_6$  grand unified theory the exotic $D, D^c$ partners of the
Higgs doublets would have diquark Yukawa couplings such as
$W_{DQ}\sim DQQ$ or $D^cu^cd^c$, as well as leptoquark couplings
$W_{LQ}\sim Du^ce^c$ or $D^cQL$, which are related by $E_6$
to the ordinary Higgs Yukawa couplings. These would lead to rapid proton decay mediated
by the $D$ and $D^c$ unless their masses (and therefore the \uprm\ breaking scale)
is comparable to the unification scale. A TeV-scale \zpr\ therefore requires that the
GUT Yukawa relations are not respected, so that either the leptoquark or the diquark
couplings (or both) are absent. This could come about in a string
construction if the fields in the multiplet are not directly related to each other in the
underlying theory (see, e.g.,~\cite{Witten:2001bf}).
See~\cite{King:2005jy,Howl:2007zi} for a detailed study of 
 complete $E_6$ models with a low energy \uprm.
Alternatively, one can simply view the charges and exotics as an example
of an anomaly-free construction.

\begin{table*}[ht]
\caption{Decomposition of the $E_6$ fundamental representation
of left-handed fermions
${\bf 27}$ under $SO(10)$ and $SU(5)$, and their $U(1)_{\chi}$,
$U(1)_{\psi}$, $U(1)_{\eta}$, inert $U(1)_I$,
neutral-$N$ $U(1)_N$, and  secluded sector $U(1)_S$ charges.
A general model in this class has
charge $Q_2=\cos\theta_{E_6}
Q_\chi+\sin\theta_{E_6} Q_\psi-\epsilon Y$,
where $\epsilon$ can result from kinetic mixing,
and coupling $g_2=\sqrt{\frac{5}{3}} g \tan \theta_W \lambda_g^{1/2}$,
where $ \lambda_g$ is usually of ${\cal O} (1)$. \label{tab:e6decomp}}
\begin{center}
\begin{tabular}{|c| c| c| c| c|c|c|c|}
\hline $SO(10)$ & $SU(5)$ & $2 \sqrt{10} Q_{\chi}$ & $2 \sqrt{6}
Q_{\psi}$ & $2 \sqrt{15} Q_{\eta}$  & $2Q_I$ &$2 \sqrt{10} Q_N$& $2 \sqrt{15} Q_S$\\
\hline
16   &   $10~ (u,d,{u^c}, {e^+} )$ & $-$1 & 1  & $-$2&0  &1& $-{1/2}$\\
            &   ${5^\ast}~ ( d^c, \nu ,e^-)$  & 3  & 1 & 1&$-1$  & 2  & 4  \\
            &   $\nu^c$             & $-5$ & 1  & $-5$  &1  & 0    & $-5$  \\
\hline
       10   &   $5~(D,H_u)$    & 2  & $-2$ & 4   &0  & $-$2    & 1  \\
            &   ${5^\ast} ~(D^c, H_d)$ & $-2$ &$-2$ & 1 &1& $-$3  & $-{7/2}$\\
\hline
       1    &   $1~ S$                  &  0 & 4 & $-5$&$-1$ & 5 & $5/2$\\
\hline
\end{tabular}
\end{center}
\end{table*}

\subsection{Anomaly-Free Sets}\label{sec:anomfree}

Many authors have described classes of \uprm\ models by requiring the
cancellation of anomalies and other criteria~\cite{Appelquist:2002mw,Cvetic:1997ky,Barr:1986hj,Cheng:1998nb,Cheng:1998hc,Erler:2000wu,Joshipura:2000sn,Ma:2002tc,Carena:2004xs,Demir:2005ti,Batra:2005rh,Morrissey:2005uz,Kang:2007ib,Lee:2007fw,Langacker:2007ac}.
Usually some conditions are applied on the types of exotics. 
It is usually assumed that any exotic fermions are non-chiral under the standard model,
i.e., that they occur in vector pairs $\psi+\psi^c$. This avoids the introduction of any SM anomalies
and also reduces the sensitivity to precision electroweak constraints~\cite{Yao:2006px}.
One can then constrain the exotic  representations with respect to the SM and their \uprm\
charges from the mixed SM-\uprm\ conditions
\beq \sum_{f\in 3,3^*} Q_{2f}= \sum_{f\in2} Q_{2f}=\sum_fY^2Q_{2f}=\sum_f Y Q_{2f}^2=0. 
\eeql{mixedanom}
The pure \uprm\ conditions $\sum_f  Q_{2f}=\sum_f  Q_{2f}^3=0$ further restrict the charges.
Alternatively, some authors ignore the latter because they can be satisfied by adding
SM singlets to the model. This can always be done with rational charges\footnote{One expects
the charges to be rational if the \uprm\ is embedded in a simple group, but this need 
not be the case for more complicated embeddings, such as the SM couplings in {Eq.~\ref{SMcouplings}}.} if
the mixed anomaly solutions are rational~\cite{Batra:2005rh,Morrissey:2005uz}.
However, because of its cubic nature the singlet structure is sometimes complicated.

In non-supersymmetric models it is often assumed that the only chiral fermions
are the three ordinary families and 3 corresponding families of  exotics.
This assumption is often not valid in supersymmetric models, where one must
also take into account the fermionic partners of the Higgs doublets and of the SM singlets
which break the \uprm. These, as well as other exotics, often do not occur in three families
(exceptions are the  $E_6$ models, where they do occur in
three families, and the $T_{3R}, B - L$ models, where the Higgs doublets and singlets are usually non-chiral).

Another issue in the supersymmetric models is whether the MSSM unification of the
SM gauge couplings is preserved. The simplest way for this to occur is for the
three SM families, which transform as ${\bf 10} + {\bf 5^\ast}$ under $SU(5)$,
and the two Higgs doublets $H_{u,d}$,
are supplemented by exotics which transform as  ${\bf 5} + {\bf 5^\ast}$  and/or  ${\bf 10} + {\bf 10^\ast}$.
It is {\em not} necessary for the fields in a $SU(5)$ multiplet to have the same \uprm\
charges (e.g., they may have different origins in an underlying string theory), and in fact
under minimal assumptions they must be different~\cite{Morrissey:2005uz}. 
An alternative is to allow non-chiral exotics, as in the $E_6$ models, 
reintroducing a form of the $\mu$ problem.

Other conditions are often employed along with the anomaly and unification constraints. These may involve the
existence of quark and lepton Yukawa couplings for one or two Higgs doublets, constraints on neutrino mass, Yukawas that can lead to masses for the exotics, operators that can allow exotic decays,
whether the charges are family universal,
whether the \uprm\ solves the supersymmetric $\mu$ problem,
whether it forbids $R$-parity violating operators~\cite{Erler:2000wu,Joshipura:2000sn,Ma:2002tc}
 or other operators relevant to proton decay~\cite{Lee:2007fw,Chamseddine:1995rs,Coriano:2007ba,Lee:2007qx}, 
whether it plays the role of a family symmetry relevant to the fermion masses
and mixings~\cite{Joshipura:2000sn,Kaplan:1998jk}, 
and many other possible conditions.

The  $Q^{YBL}$ models of Section~\ref{lrmodels}, which do not require any exotics other than $\nu_R$,
are discussed in~\cite {Appelquist:2002mw}. Four one-parameter
families of models with three families of exotics were constructed in~\cite{Carena:2004xs}.
Two of these, referred to as $q+xu$ and ${\bf 10}+x {\bf 5^\ast}$ are equivalent to the
$Q^{YBL}$ and the $E_6$ model $Q(\theta_{E_6})$, respectively, while the others
($B-xL$ and $d-xu$),
have not emerged from other considerations for general $x$. The ${\bf 10}+x {\bf 5^\ast}$ and 
$d-xu$ would require two Higgs doublets to have normal quark and charged lepton Yukawas.

The most systematic classification of the supersymmetric models is given in~\cite{Erler:2000wu},
which requires anomaly cancellation, minimal gauge unification with no non-chiral states,
exotic masses, 
and the absence of rapid proton decay or fractional electric charges. Classes of solutions were found,
which required that more than one SM singlet participates in the \uprm\ breaking.
A particularly simple one is the $Q_{\tilde \psi}$ model. It involves two
${\bf 5} + {\bf 5^\ast}$ pairs $(D_i+L_i)$ and $(D^c_i+L^c_i), i=1,2$, 
which are analogous to the $(D,H_u)$ and $(D^c, H_d)$ of the $E_6$ model,
along with $H_{u,d}$ and the three SM families. The \uprm\ symmetry is broken by the VEVs of two singlets, $S$ and  $S_D$, which also generate masses for the $H_{u,d}$ and $L_i,L^c_i$ ($\langle S \rangle$) and for $D_i, D^c_i$ ($\langle S_D \rangle$). Additional singlets are needed for
the \uprm\ anomalies. The $Q_{\tilde \psi}$ charges are listed in Table~\ref{tab:gaugeunif}.
The fermion currents are purely axial.
It is straightforward to generalize the  $Q_{\tilde \psi}$ model to $Q_{\bf 55^\ast}$, which allows non-axial charges
and $n_{\bf 55^\ast}$ pairs of  ${\bf 5} + {\bf 5^\ast}$. Three distinct chiral singlets 
must acquire VEVs to generate all of the exotic masses, except for $n_{\bf 55^\ast}=2$ or $3$.
Additional SM singlets are needed for
the \uprm\ anomalies and to generate singlet masses.
The gauge coupling $g_2$ is arbitrary. 

It was shown in~\cite{Demir:2005ti} that anomaly-free supersymmetric models
can be constructed without any exotics (not even $\nu^c$)
and only one singlet $S$ (which generates a dynamical $\mu_{eff}$) provided one allows  family nonuniversal charges (an early example was also given in~\cite{Cvetic:1997ky}).
It is possible to choose the charges to avoid flavor changing neutral current (FCNC)
effects (see Section~\ref{sec:fcnc}). However, the \uprm\ forbids some of the quark and lepton Yukawa interactions in the
superpotential. 
These could possibly be generated by non-holomorphic soft terms, as described in Section~\ref{twodoublet}.

\begin{table}[ht]
\caption{Examples of supersymmetric models consistent with minimal SM gauge
unification. $n_{\bf 55^\ast}$ is the number of pairs of  ${\bf 5} + {\bf 5^\ast}$.
$Q_S$ is taken to be 1. The free parameters are $Q_{H_u}\equiv x, Q_Q\equiv y,
Q_D\equiv z$ (which only affects the exotics), and 
the gauge coupling $g_2$. Kinetic mixing can be added. The $Q_{\tilde \psi}$
model is a special case with axial charges and  $n_{\bf 55^\ast}=2$. Additional SM
singlets are not displayed. The $\nu^c$ charge allows a Dirac $\nu$ mass term.
 \label{tab:gaugeunif}}
\begin{center}
\begin{tabular}{|c|c| c|| c| c| c|}
\hline &$Q_{\bf 55^\ast}$ &  $Q_{\tilde \psi}$& & $Q_{\bf 55^\ast}$ &  $Q_{\tilde \psi}$ \\
\hline
$Q$ & $y$ & $1/4$          & $H_u$ & $x$ & $-1/2$ \\
$u^c$ & $-x-y$ & $1/4$      & $H_d$ & $-1-x$ & $-1/2$ \\
$d^c$ & $1+x-y$ & $1/4$      & $S_D$ & $3/n_{\bf 55^\ast}$ & $3/2$ \\
$L$ & $1-3y$ & $1/4$          & $D_i$ & $z$ & $-3/4$ \\
$e^+$ & $x+3y$ & $1/4$      & $D_i^c$ & $-3/n_{\bf 55^\ast}-z$ & $-3/4$ \\
$\nu^c$ & $-1-x+3y$ & $1/4$     & $S_L$ & $2/n_{\bf 55^\ast}$ & $1$ \\
$S$ & $1$ & $1$         & $L_i$ &  $\frac{5-n_{\bf 55^\ast}}{4n_{\bf 55^\ast}}+x+3y+3z/2$ & $-1/2$ \\
&&      & $L_i^c$ & $-2/n_{\bf 55^\ast}-Q_{L_i}$ & $-1/2$ \\
\hline
\end{tabular}
\end{center}
\end{table}

\subsection{TeV Scale Physics Models}\label{sec:tev}
In this section we briefly consider various models involving new TeV scale
physics, especially those motivated as alternatives to the elementary Higgs
for electroweak symmetry breaking. 

As a preliminary, consider a direct product of two identical gauge group $G\equiv G_1\x G_2$,
with  generators  $\vec T_{1,2}$ and associated currents $\vec J_{1,2}$. Then

\beq
-L=g_1 \vec J_1 \cdot \vec W_1+ g_2 \vec J_2 \cdot \vec W_2. \eeql{prod}
$G$ can be spontaneously broken to the {\em diagonal subgroup} $G_D$ with
generators $\vec T_D=\vec T_1+\vec T_2$ if there is a Higgs field which transforms
equivalently under both groups.
An example is $SU(N)\x SU(N)$, with a Higgs $\varphi^\alpha_a$ transforming as ${\bf N^\ast}\x
{\bf N}$, with $\langle \varphi^\alpha_a \rangle=  c \delta^\alpha_a$. 
It is then
straightforward to show that 
\beq
-L=g_L( \vec J_1+ \vec J_2) \cdot \vec W_L+ g_L(\cot \delta \vec  J_1-\tan \delta \vec J_2 )\cdot \vec W_H, \eeql{proddiag}
where $\vec W_L= \sin \delta \vec W_1 + \cos \delta \vec W_2$ is the massless boson, $W_H$
is the massive orthogonal combination, $\tan \delta = g_2/g_1$, and $g_L=g_1 \sin \delta$.
$W_L$ can acquire mass and $W_{L,H}$ can mix due to additional Higgs fields.
A simple illustration  is the SM breaking of $U(1)_{T_3}\x U(1)_Y$
to $U(1)_{em}$ by the ordinary Higgs doublet.

\subsubsection{Little Higgs, Twin Higgs, and Un-Unified Models}\label{sec:littlehiggs}
In {\em Little Higgs} models~\cite{ArkaniHamed:2001nc} the Higgs is a pseudo-Goldstone
boson of an approximate global symmetry. (For reviews, see~\cite{Han:2003wu,Han:2005ru,Perelstein:2005ka}.) The one-loop (and sometimes two-loop) quadratic divergences in the
Higgs mass-square are cancelled by new TeV gauge bosons, fermions, and
scalar particles related to those of the SM. There are a wide class of models, all of which involve
heavy neutral and charged gauge bosons. For example, in the Littlest Higgs
models~\cite{ArkaniHamed:2002qy} the electroweak gauge group is
$[SU(2)\x U(1)]^2$, which is a subgroup of a larger global group. The SM
left-handed fermions are charged under only the first $SU(2)$. The $SU(2)^2$ symmetry
is broken by a condensate charged under both factors to an unbroken diagonal
subgroup, 
and the $U(1)$ 
charges are chosen to yield $U(1)_Y\x U(1)_H$, where $Y$ is the normal hypercharge.
Thus, the residual gauge group is $SU(2)_L\x U(1)_Y \x SU(2)_H\x  U(1)_H$.
From {Eq.~\ref{proddiag}}, 
the heavy charged $W^\pm_H$ and neutral $W^0_H$ couple to the left-handed 
SM quarks and leptons with the $SU(2)_L$ generators ${\vec \tau/2} $
and with coupling $g\cot\delta$. The neutral $U(1)_H$ boson
is lighter, with model dependent couplings.  Precision electroweak constraints
are rather severe, unless one pushes the Little
Higgs scale to be uncomfortably large compared to the original motivation. 
However, the difficulties can be reduced if the $U(1)_H$ is not gauged.

The precision electroweak constraints are greatly weakened
(they are only
generated at loop level) if one introduces a
discrete symmetry, $T$-parity~\cite{Cheng:2003ju,Hubisz:2005tx}. This is analogous to $R$-parity in supersymmetry, and requires that the heavy states, such as the new gauge bosons,
only couple in pairs to the ordinary particles. This also means that they must be pair produced
at colliders. The lightest could be stable and possibly be a dark matter candidate.
However, it has recently been argued that the $T$-parity may be broken by anomalies~\cite{Hill:2007zv}, leading to decays, e.g., into $ZZ$.

In the {\em Twin Higgs} model~\cite{Chacko:2005pe} the Higgs quadratic divergences
are canceled by particles from a hidden sector that is a mirror of the SM and which mainly communicates
by an extended Higgs sector. The gauge bosons in the hidden sector may essentially decouple
from the SM particles and could even be massless, while in other versions there may be kinetic
mixing with the photon. 

In the {\em Un-unified} model~\cite{Georgi:1989xz}, the left-chiral SM quarks and leptons
transform under distinct $SU(2)$ groups $SU(2)_q$ and $SU(2)_l$ with gauge
couplings $g_{q,l}$, i.e., they are not unified. There is a single conventional $U(1)_Y$.
After diagonal breaking, one recovers the SM along with heavy $W^{\pm,0}_H$
which couple to $g(\cot \delta \vec  J_q-\tan \delta \vec J_l )$ using  {Eq.~\ref{proddiag}}.
For small $\tan \delta $ the heavy bosons couple mainly to quarks.

\subsubsection{Extra Dimensions}\label{sec:extradim}
The existence of extra dimensions is suggested by string models~\cite{Antoniadis:1990ew}.
There are a wide variety of models, depending on
their number, size, whether they are flat or warped, whether the SM fields are allowed
to propagate in the extra dimensions (i.e., in the {\em bulk}), etc. 
For a review, see~\cite{Yao:2006px}.

The simplest case involves a single extra dimension of radius $R$, implying
the existence of Kaluza-Klein excitations of the states that can propagate in the bulk,
with mass $\sim n/R, n=1, \cdots$. If only gravitons 
propagate, then $R$ can be large enough to probe in laboratory
gravity experiments. However, if the SM gauge bosons are also allowed to
propagate, then $R^{-1}$ must be larger than  ${\cal O}$(TeV) ($R\lesssim 10^{-17}$ cm).
If the SM fermions and Higgs are not allowed to propagate (i.e., {\em confined to the brane}),
then the excitations of the SM gauge bosons ($W^\pm, Z, A,$ gluon) couple to the same
currents as their SM counterparts, but with a coupling constant larger by $\sqrt 2$~\cite{Masip:1999mk,Casalbuoni:1999ns}. Current experimental limits require $R^{-1} \gtrsim 7$ TeV~\cite{Cheung:2001mq,Barbieri:2004qk}. The limits are much weaker ($\cal O$(300 GeV)) in {\em universal extra dimension} models, in which
all of the SM fields propagate uniformly in the bulk~\cite{Appelquist:2000nn,Cheng:2002ab,Appelquist:2002wb,Gogoladze:2006br}.
Similar to $R$ or $T$-parity, there is a $KK$-parity so that the $n=1$ states 
can only be pair produced and only contribute to electroweak observables in loops. The lightest
is stable. In variants in which the various quarks and leptons are localized in different parts of the
extra-dimensional space (with implications for the flavor problem) the couplings of the
Kaluza-Klein excitations are family nonuniversal (since the overlap of the wave functions
depends on location). This leads to the possibility of FCNC  effects~\cite{Delgado:1999sv},
as discussed in Section~\ref{sec:fcnc}.

Models involving warped extra dimensions~\cite{Randall:1999ee} may have all of the
SM fields confined to the infrared brane. However, much attention has been devoted to
the possibility that the SM fields other than the Higgs can also propagate in the bulk~\cite{Hewett:2002fe,Carena:2003fx,Agashe:2003zs,Agashe:2007ki}, e.g.,
because in that case the theory is related to technicolor models by the
AdS/CFT correspondence~\cite{Maldacena:1997re}.
It is then useful to enhance the electroweak gauge symmetry to
$SU(2)_L\x SU(2)_R \x U(1)_{B-L}$ to provide a custodial symmetry to
protect the electroweak $\rho$ parameter~\cite{Agashe:2003zs}. The Kaluza-Klein
excitations of the gauge bosons couple mainly to the $t$ and $b$ due to wave
function overlaps, and decays to $WW$ and $Z+$ Higgs are also possible due to mixings (see, e.g.,~\cite{Agashe:2007ki}).

\subsubsection{Strong Dynamics}\label{sec:strong}
There have been many models in which strong dynamics is involved in electroweak
symmetry breaking, which often involve additional elementary gauge bosons or
composite spin-1 states, which may be strongly coupled.

Dynamical symmetry breaking (DSB) models in which the Higgs is replaced by a fermion condensate
are reviewed in~\cite{Hill:2002ap,Chivukula:2002ry,Chivukula:2003wj}. For example, {\em topcolor} models~\cite{Hill:1994hp}
typically involve new gluons and a new \zpr\ that couple preferentially 
and with enhanced strength to the third generation
and which assist in forming a top condensate. Nonuniversal extended technicolor models~\cite{Chivukula:2002ry}
also feature new gauge interactions preferentially coupled to the third family. 

The {\em BESS} (Breaking Electroweak Symmetry Strongly) models~\cite{Casalbuoni:1985kq,Casalbuoni:1986vq} are effective
Lagrangian descriptions of models with a strongly interacting longitudinal  gauge boson sector, such as one expects in the large $M_H$ limit of the SM or in some forms of DSB. There are vector and
axial bound states which can mix with the $W^\pm, Z$, and $A$. They interact with the SM particles
directly and by mixing.
The possibility that the  electroweak bosons could be composite has also been considered~\cite{Baur:1985ez}. 

Another interesting  model with no elementary or composite Higgs~\cite{Csaki:2003zu} is a variant on the warped extra dimension scenario. However, instead of including a Higgs field the electroweak symmetry is broken by boundary conditions. The Kaluza-Klein excitations of the gauge bosons unitarize the high energy scattering of longitudinal gauge bosons. More general classes of Higgless models
may involve fermiophobic \zpr\ which may be produced and detected by their couplings to the $W$ and
$Z$~\cite{He:2007ge}.

\subsection{Non-Standard Couplings}\label{sec:modified}
Most of the canonical \zpr\ models assume electroweak scale couplings, and
that the \zpr\ couplings to most or all of the SM fermions are of comparable strength
and family universal,
in which case existing experimental constraints require masses not too much
below 1 TeV (Section~\ref{sec:exp}). However, there are many models with different
assumptions concerning the gauge couplings, charges, and scales.

\subsubsection{ Decoupled Models}\label{sec:reduced}
{\em Leptophobic} \zpr~s~\cite{delAguila:1986iw} do not couple to ordinary neutrinos
or charged leptons, and therefore most direct electroweak and collider searches
are insensitive to them. They could emerge, e.g., in the $E_6$ $\eta$ model
in Table~\ref{tab:e6decomp}, combined with a (large) kinetic mixing~\cite{Babu:1996vt}
 $\epsilon\sim -1/\sqrt{15}$; in a flipped $SU(5)$ model~\cite{Lopez:1996ta}; or in models in which the \zpr\ couples
to baryon number~\cite{Carone:1994aa}. Approximately leptophobic models were once
suggested by apparent anomalies in $Z\ra b \bar b$ decays (see~\cite{Rosner:1996eb,Umeda:1998nq} for references), but are still an interesting possibility for allowing \zpr\ masses much smaller than a TeV. A purely leptophobic \zpr\ is still constrained by $Z-\zpr$ mixing effects~\cite{Umeda:1998nq},
and could be inferred by collider signals such as the production of $t\bar t$ pairs, exotics~\cite{Rosner:1996eb}, or the same-sign dilepton decays of a pair of heavy Majorana neutrinos~\cite{Duncan:1986wz,delAguila:2007ua}.
They could even be light enough to be produced in $\Upsilon$ decays~\cite{Aranda:1998fr}.

Limits are also weak, e.g., if the \zpr\ couples only to the second and third family
leptons~\cite{Foot:1994vd} or if it couples only to third family fermions~\cite{Andrianov:1998hx}. 
In {\em fermiophobic} models~\cite{Barger:1980ix,Donini:1997yu}
there are no direct couplings of the \zpr\ to the SM fermions, although they may be induced by ordinary 
or kinetic mixing. An interesting possibility is that such fermiophobic \zpr\ may couple to
a hidden sector~\cite{Kumar:2006gm,Chang:2006fp}, such as may be associated with supersymmetry breaking. Mixing effects could therefore
possibly be a means of probing such a sector (direct \zpr\ couplings to a hidden sector are
considered in Section~\ref{sec:hidden}).
Finally, a \zpr\ with canonical charges could still be much lighter than a 
TeV if its gauge coupling is sufficiently small~\cite{Fayet:1980ss,Freitas:2004hq,Fayet:2007ua,Nelson:2007yq}.

\subsubsection{St\"uckelberg Models}\label{sec:stueckelberg}
It is possible to write a $U(1)$ gauge invariant theory with a massive gauge boson $C_\mu$ 
by the St\"uckelberg mechanism~\cite{Stueckelberg}. 
The Lagrangian is
\beq
L=-\frac{1}{4} C^{\mu\nu}C_{\mu\nu}-\oh (mC^\mu+\partial^\mu \sigma) (mC_\mu+\partial_\mu \sigma),
\eeql{stueckelberglag}
where $C_{\mu\nu}$ is the field strength tensor. Under a gauge
transformation, $\Delta C_\mu = \partial_\mu \beta$, while the field $\sigma$
is shifted,  $\Delta \sigma = -m \beta$, analogous to the shift in an axion field.
A gauge-fixing term can be added to {Eq.~\ref{stueckelberglag}} which cancels the cross
term between $C$ and $\sigma$, leaving a massive $C$ field and a decoupled $\sigma$.
This is analogous to the Higgs mechanism, but there is no field with a VEV and no
physical Higgs boson. This mechanism has recently been applied to a \uprm\
extension of the SM or the MSSM~\cite{Kors:2004dx,Feldman:2007nf}.
For example, if one replaces the second term in {Eq.~\ref{stueckelberglag}} by
$-(M_2 C_\mu+M_1 B_\mu+\partial_\mu \sigma)^2/2$ with $M_1/M_2\equiv \epsilon \ll1$,
then the $C$ will mix with the $A$ and $Z$, but there will remain a massless  photon.
The new $Z_2$ can be relatively light (e.g., several hundred GeV), so $\epsilon$ must be
small. If
the $C$ has no direct couplings to matter, the new $Z_2$ will decay only to SM particles via the mixing
and will be very narrow. If the $C$ does couple to exotic matter, then the mixing with the photon
will induce tiny (generally) irrational electric charges of ${\cal O}(\epsilon)$ for the exotic particles.
Such mixing with the photon is never induced by ordinary Higgs-type mixing 
if $U(1)_{em}$ is unbroken, but can also be induced by kinetic mixing with another massless boson
(Section~\ref{sec:massless}).
Other applications, such as to dark matter, are reviewed in~\cite{Feldman:2007nf}.

\subsubsection{Family Nonuniversal Models}\label{sec:nonuniversal}
Another variant is the possibility of {\em family nonuniversal} charges (e.g.,~\cite{Demir:2005ti}). A number of examples
of \zpr\ coupling preferentially to the third family or to the $t$ quark were mentioned in Sections \ref{sec:extradim},
\ref{sec:strong}, and \ref{sec:reduced}. 
These could have enhanced gauge couplings, and could be observed
as a resonance in $t\bar t$ production. String-derived \zpr~s often have nonuniversal couplings
as well (Section~\ref{sec:anomalous}), as do the Kaluza-Klein excitations
in extra-dimensional theories in which the fermion families are spatially separated
(Section~\ref{sec:extradim}). Possible FCNC effects are considered in Section~\ref{sec:fcnc}.

\subsection{\uprm\ Breaking Scales}\label{sec:massscales}
Most attention is given to possible electroweak or TeV scale \zpr~s, but there are other possibilities.
Here we describe massless, TeV scale, and intermediate scale models.
Models involving the GUT or string scales are described in Section~\ref{sec:anomalous}.

\subsubsection{A Massless \zpr}\label{sec:massless}
A \zpr\ would be massless if the \uprm\ symmetry is unbroken. This would imply an unacceptable
long range force if it coupled to ordinary matter unless the coupling were incredibly small~\cite{Dobrescu:2006au}. It would be allowed if the primary coupling were to a hidden sector and
communicated only by higher-dimensional operators~\cite{Dobrescu:2004wz} 
or by kinetic mixing with the photon~\cite{Holdom:1985ag}. The latter scenario would induce a 
small fractional electric charge for hidden sector particles.

\subsubsection{Electroweak/TeV Scale \zpr}\label{sec:tevscale}
Models in which the \uprm\ is involved in electroweak symmetry breaking,
such as in Section~\ref{sec:tev}, typically involve \uprm\ breaking
at the electroweak or TeV scale. 

In the \uprm\ extension of the MSSM
with a single $S$ field~\cite{Cvetic:1997wu,Cvetic:1995rj,Cvetic:1997ky,Keith:1997zb,Langacker:1998tc}, the part of the superpotential involving
$S$ and $ H_{u,d}$ is $W= \lambda_S S H_u H_d$,
where we have assumed $Q_S\ne0$ and $Q_S+Q_u+Q_d=0$. Like the MSSM, the minimum
of the tree-level potential always occurs along the charge-conserving direction
with only  $\langle H^0_{u,d} \rangle\ne 0$
(this assumes that the squark and slepton VEVs vanish). The potential is then
\beq V= V_F+V_D+V_{soft}, \eeql{umssmpotential}
where
\beq \begin{split}
V_F & =  \lambda_S^2 \left( |H_u^0|^2 |H_d^0|^2 + |S|^2 |H_u^0|^2 + |S|^2 |H_d^0|^2 \right) \\ 
V_D & =  \frac{g_1^2}{8} \left( |H_u^0|^2 - |H_d^0|^2 \right)^2 \\
& +
\frac{g_2^2}{2} \left( Q_u |H_u^0|^2 + Q_d |H_d^0|^2 +Q_S |S|^2 \right)^2 
 \\ 
V_{soft} & =  m_u^2 |H_u^0|^2 + m_d^2 |H_d^0|^2 +m_S^2 |S|^2\\
&- \left( \lambda_S A_S S H_u^0 H_d^0 + {\rm h.c.} \right).
\end{split} \eeql{MSSMUV}
If $S$ acquires a VEV, then the effective $\mu$ parameter is $\mu_{eff}= \lambda_S \langle S \rangle$,
the corresponding effective $B\mu$ is $(B\mu)_{eff}= \lambda_S A_S  \langle S \rangle$, and
the $Z-\zpr$ mass matrix is given by {Eq.~\ref{massmatrix}} and {Eq.~\ref{examplepars}}.
One can define the fields so that $ \lambda_S A_S $ and therefore the VEVs 
$\nu_{u,d}$ and $s$ defined after {Eq.~\ref{examplepars}} are real and positive.
There is no analog of the first (second) term in $V_F$ ($V_D$) in the MSSM.

For generic parameters one expects $\nu_{u,d}$ and $s$ to be comparable.
For example, for $ \lambda_S A_S $ large compared to the soft masses and
$Q_u=Q_d=-Q_S/2$ one finds~\cite{Cvetic:1997ky} $\nu_u\sim \nu_d \sim s$, with negligible $Z-\zpr$
mixing and $\mzp^2/M_Z^2\sim 12 g_2^2 Q_u^2/g_1^2$, which is typically of order 1.
This case is excluded unless the model is leptophobic or something similar.

A more likely scenario is that the soft parameters 
 ($|m_{u,d,S}|, |A_S|$) are of $\cal O$(1~TeV), with
$m_S^2 < 0$. Then $s^2 \sim -2 m_S^2/g_2^2 Q_S^2$ and $\mzp^2 \sim
-2 m_S^2$. One can have a smaller EW scale $\nu_{u,d} \ll s$ by accidental
cancellations, which are not excessive provided \mzp\ is not too much larger than a TeV.
In most supersymmetry mediation schemes $m_S^2$ is positive at a large scale
such as the Planck scale. The running $m_S^2$ can be driven negative at low scales 
radiatively provided it has sufficiently large Yukawa couplings, such as $\lambda_S$ 
and/or couplings to exotics such as in {Eq.~\ref{super}}. This is analogous to the MSSM in which
$m_u$ can be driven negative by its large Yukawa coupling to the top.

\subsubsection{Secluded Sector and Intermediate Scales}\label{sec:intermediate}
In the single $S$ model in {Eq.~\ref{umssmpotential}} and (\ref{MSSMUV}) there is some tension between the electroweak
scale and developing a large enough \mzp.
These can be decoupled without tuning when there are several $S$ fields.
 For example, in the {\em secluded sector}
model~\cite{Erler:2002pr} there are four standard model singlets $S, S_{1,2,3}$ that are
charged under a \uprm, with
\beq
W= \lambda_S S H_u H_d + \lambda S_1 S_2 S_3.
\eeql{wsecluded}
(Structures similar to this are often encountered in heterotic string constructions.)
$\mu_{eff}$  is given by $\lambda_S \langle S \rangle$, but all four VEVs contribute
to \mzp. The only couplings between the ordinary $(S, H_{u,d})$
 and secluded $(S_{1,2,3})$ sectors are from the \uprm\ $D$ term and
 the soft masses (special values of the \uprm\ charges, which allow soft mixing terms, are required
 to avoid unwanted additional global symmetries). It is straightforward to choose the
 soft parameters so that there is a runaway direction in the limit $\lambda\ra 0$,
 for which the ordinary sector VEVs remain finite while the $S_i$ VEVs become large.
 For $\lambda$ finite but small, the $\langle S_i \rangle$ and \mzp\ scale as 
 $1/\lambda$. For example, one can find \mzp\ in the TeV range for $\lambda\sim 0.05-0.1$.
 The secluded model can be embedded in the $E_6$ context (Table~\ref{tab:e6decomp}).
 
 {\em Intermediate Scale} models~\cite{Cleaver:1997nj,Morrissey:2006xn} are those in which the \uprm\ breaking is associated with
 a $F$ and $D$ flat direction, such as the secluded model in {Eq.~\ref{wsecluded}} with $\lambda=0$.
 However, let us consider a simpler toy model with two fields $S_{1,2}$ with $Q_{S_1} Q_{S_2}<0$.
 If there are no terms in $W$ like $S_iS_j$ or $S_iS_jS_k$, then the potential for
 $S_{1,2}$ is
 \beq
\begin{split}
 V(S_1,S_2) &= m_1^2 |S_1|^2 + m_2^2 |S_2|^2 \\
 &+ \frac{g_2^2}{2}
( Q_{S_1} |S_1|^2 + Q_{S_2} |S_2|^2)^2.
\end{split}
 \eeql{intscale}
 The quartic term vanishes for $|S_2|^2/|S_1|^2
= - Q_{S_1}/Q_{S_2}$. For simplicity, take 
$Q_{S_1}= -Q_{S_2}$, and assume that at low energies $m_{S_1}^2 < 0$
and $m_{S_2}^2 > 0$, as would typically occur by the
radiative mechanism if $W$ contains a term 
$h_D {S_1} {D}  {D^c}$. If $m^2 \equiv 
m_{S_1}^2 + m_{S_2}^2 >0$ the minimum will occur at
$\langle S_1 \rangle \ne 0, \ \langle S_2 \rangle = 0$.
If there is also a $\lambda_S S_1 H_u H_d$ term in $W$ then
$\langle S_1 \rangle$
and $M_{Z'}$ will be at the EW scale ($\lesssim$ 1 TeV), just as
in the case of a single $S$. 
On the other hand, for $m^2 < 0$,
the potential along the $F$ and $D$ flat direction
$S_1 = S_2 \equiv S$ is
\begin{equation}
V(S) = m^2 S^2,
\label{unbounded}
\end{equation}
which appears to be unbounded from below. In fact, $V(S)$ is  typically stabilized by one or 
both of two mechanisms: (a) The leading loop corrections to the effective (RGE-improved)
potential
result in $m^2 \rightarrow m^2(S)$, leading to a minimum
slightly below the scale at which $m^2(S)$ goes through zero, 
which can be  anywhere in the range $10^3 - 10^{17}$ GeV. (b) Another possibility is that 
the $F$-flatness is lifted by higher-dimensional
operators (HDO) in $W$,  such as $W = ({S}_1 {S}_2)^k/M^{2k-3}$,
where $M$ is the Planck or some other large scale. This would lead to $\langle S \rangle\sim
\sqrt{mM}\sim 10^{11}$
GeV for $k=2$, $m\sim 1$ TeV, and $M$ the Planck scale. In such models, HDO
such as $LH_de^+ (S/M)^p$ or $S H_u H_d (S/M)^q$ could also be important for generating
small effective Yukawa couplings  (and therefore fermion mass hierarchies)
or $\mu_{eff}\ll \langle S \rangle$ terms~\cite{Cleaver:1997nj}.
Implications for neutrino mass are considered in Section~\ref{sec:neutrino}.

\subsection{Grand Unification, Strings, and Anomalous \uprm}\label{sec:anomalous}

\subsubsection{Grand Unification}\label{sec:gut}
In some full grand unified theories~\cite{Langacker:1980js,Hewett:1988xc},  such as  $E_6$, the extra \uprm~s must be broken
at or near the GUT unification scale to avoid rapid proton decay. 
This typically occurs if the Higgs doublets (and their Yukawa couplings to
ordinary fermions) are related by the GUT symmetry to chiral exotics, which cannot be much heavier than
the \uprm\ breaking scale. However, as mentioned in
Section~\ref{sec:e6model} this can be evaded in models which respect the GUT quantum numbers but
not the Yukawa relations, or in models, such as the $E_6$ $\chi$ model, in which the Higgs doublets are non-chiral.

\subsubsection{String Theories}\label{sec:string}

Most semi-realistic superstring constructions yield effective four-dimensional field theories
that include the SM gauge group ({\em not}  a full four-dimensional GUT), as well as additional
gauge group factors that often 
involve additional \uprm~s.
(Examples include~\cite{Faraggi:1990ita,Faraggi:1991mu,Cleaver:1998gc,Giedt:2000bi,Cvetic:2001nr,Braun:2005bw,Anastasopoulos:2006da,Lebedev:2007hv,Coriano:2007ba}. For  reviews, see~\cite{Cvetic:1997wu,Blumenhagen:2005mu,Blumenhagen:2006ci}.)
Heterotic constructions often descend through an underlying $SO(10)$ or $E_6$ in the higher-dimensional
space, and may therefore lead to the $T_{3R}$ and $B - L$ (i.e., $Q^{LR}$) or the $E_6$-type charges.
Additional or alternative  \uprm\ structures may emerge that do not have any GUT-type interpretation
and therefore  have very model dependent charges.
Similarly, intersecting brane constructions often descend through Pati-Salam type
models~\cite{Mohapatra:737303}, yielding $Q^{LR}$. Other branes can lead to
other types of \uprm\ charges. For example, the construction in~\cite{Cvetic:2001nr}
involves two extra \uprm~s, one coupling to  $Q^{LR}$ and the other only
to the Higgs and the right-handed fermions.

Constructions often have one or multiple SM singlets which can acquire VEVs to break the
extra \uprm. However, that is not always the case. For example, in some of the $Q^{LR}$ models
(see, e.g.,~\cite{Cvetic:2001nr,Braun:2005bw}) 
 the only fields available to break the enhanced gauge symmetry
are the scalar partners $\tilde{\nu}_R$ of the right-handed neutrinos~\cite{Cvetic:2002qa}.
These act like the $\delta_R^0$ defined in Section~\ref{lrmodels}, but 
 it is difficult to reconcile the \zpr\ constraints with neutrino phenomenology.
This also occurs in  the simpler supersymmetric  versions of the $\chi$
model.

The \uprm\ in string constructions may couple to hidden sector particles, and in
some cases they can communicate between the ordinary and hidden sectors~\cite{Langacker:2007ac,Verlinde:2007qk}.
The non-standard string \uprm\ often have family nonuniversal charges. This can occur if the fermion families
have different embeddings in the underlying theory. A simple field-theoretic example is a variant on the
$E_6$ model in Table~\ref{tab:e6decomp}. One could assign, e.g., the first two families
$(d^c_i, L_i), i=1,2$, to the $\bf 5^\ast$ from the $\bf 16$ of $SO(10)$, and
the third to the $\bf 5^\ast$ from the $\bf 10$.

\subsubsection{Anomalous \uprm}\label{sec:anomsection}
The effective four-dimensional field theories arising from the compactification of
a string theory usually contain anomalous \uprm\ factors
(see~\cite{Kiritsis:2003mc} for a review).
There is typically one anomalous combination in heterotic constructions.
In intersecting brane models~\cite{Blumenhagen:2005mu} there are stacks of branes yielding
$U(N)\sim SU(N)\x U(1)$, in which the $U(1)$ is usually anomalous.
Since the underlying string theory is anomaly free, these anomalies must
be cancelled by a generalized Green-Schwarz mechanism. In particular,
the \zpr\ associated with the \uprm\ acquires a  string-scale mass
by what is essentially the St\"uckelberg mechanism
in {Eq.~\ref{stueckelberglag}}, with the axion field $\sigma$ associated with an
antisymmetric field in the internal space (this sometimes applies to nonanomalous
\uprm\ as well). The \uprm\ still acts as
a global symmetry on the low energy theory, restricting the possible couplings
and having possible implications, e.g., for baryon or lepton number.
In addition, effective trilinear vertices may be generated between the \zpr\ and the SM
gauge bosons~\cite{Coriano:2005js,Anastasopoulos:2006cz}. It is possible that the string scale is actually very low (e.g., TeV scale)
if there is a large total volume of the extra-dimensional space (a realization of the
large extra dimension scenario).  This would allow TeV scale \zpr~s associated with
anomalous (or sometimes nonanomalous) \uprm~s, without any associated Higgs scalar
and with anomalous decays into $ZZ$, $WW$, and $Z\gamma$~\cite{Ghilencea:2002da,Berenstein:2006pk,Armillis:2007tb,Kumar:2007zza}.

Anomalous \uprm~s in heterotic constructions lead to Fayet-Iliopoulos (FI) terms, which are effectively
constant contributions to the \uprm\ $D$ terms that are close to the string scale.
Smaller FI terms may also appear in intersecting brane constructions which break supersymmetry.
 In many cases, FI terms trigger
scalar fields in the low energy theory to acquire VEVs to cancel them. These
VEVs in turn may lead to the breaking of gauge symmetries (such as other nonanomalous
\uprm~s) and the generation of masses for some of the particles at the FI scale,
a process known as vacuum restabilization (see~\cite{Cleaver:1998gc} for an example).
Family nonuniversal \uprm~s may be used to generate fermion textures using the Froggatt-Nielsen
mechanism~\cite{Ibanez:1994ig,Jain:1994hd,Binetruy:1994ru,Chankowski:2005qp}. The elements are associated with higher-dimensional operators
allowed by the symmetry, and involve powers of the ratio of the FI and Planck scales.

\section{Experimental Issues}\label{sec:exp}
There are limits on \zpr\ masses and $Z-\zpr$ mixing from precision electroweak
data, from direct and indirect searches at the Tevatron, and from interference effects at LEP 2.
In this section we briefly review the existing limits and future prospects for discovery
and diagnostics. FCNC effects for family nonuniversal couplings and astrophysical/cosmological
constraints are touched on in Section~\ref{sec:implications}.

\subsection{Constraints from Precision Electroweak}\label{sec:precision}
\subsubsection{Parametrization}\label{sec:precisionpar}
Precision electroweak data include purely weak $\nu e$ and $\nu$-hadron
weak neutral current (WNC) scattering; weak-electromagnetic interferences in heavy atoms
and in $e^\pm e^-$, $l^\pm$-hadron, and $\bar p p$ scattering; precision $Z$ pole physics; and associated
measurements of the $W$ and top mass. They have verified the SM at the level of radiative
corrections and strongly constrained the possibilities for new physics below the TeV scale~\cite{Yao:2006px}.
There have been a number of global analyses of the constraints from precision electroweak
on a possible \zpr~\cite{London:1986dk,delAguila:1986iw,Costa:1987qp,Durkin:1985ev,Langacker:1991pg,Chivukula:2002ry,Amaldi:1987fu,GonzalezGarcia:1990tr,delAguila:1991ry,Langacker:1991zr,Cho:1998nr,Erler:1999ub,Erler:1999nx,Cacciapaglia:2006pk}. Because of the number of different chiral fermions involved, it is difficult
to do this in a model independent way, so most studies have focussed on specific classes of
models, such as described in Section~\ref{sec:canonical}, and have emphasized electroweak scale
couplings and family universal charges.

Low energy WNC experiments are affected  by \zpr\
exchange, which is mainly sensitive to its mass, and by $Z-\zpr$ mixing.
 Prior to the Tevatron and LEP 2 they yielded the
best limits on the \zpr\ mass. The $Z$-pole experiments at LEP and SLC, on the other hand,
are mainly sensitive to $Z-\zpr$ mixing, which lowers the mass of the $Z$ relative to the SM
prediction, and also modifies the $Z\bar f f$ vertices.

The effective four-Fermi Lagrangian for the WNC obtained from {Eq.~\ref{lagrangian}} is
\beq
-L_{eff}=\frac{4G_F}{\sqrt{2}} \sum_{\alpha=1}^{n+1} \rho_\alpha 
\left[ \sum_{\beta=1}^{n+1}\frac{g_\beta}{g_1} U_{\alpha\beta} J^\mu_\beta  \right]^2,
\eeql{fourfermi}
where $\rho_\alpha\equiv M_W^2/(M^2_\alpha \cos^2 \theta_W)$, $M_\alpha$ are the mass eigenvalues, $U$ is the orthogonal transformation defined in {Eq.~\ref{eigenstates}}, and the currents are given in {Eq.~\ref{currents}} (kinetic mixing can be added). Specializing to the $n=1$ case, this is
\beq
-L_{eff}=\frac{4G_F}{\sqrt{2}} (\rho_{eff} J^2_1+2wJ_1J_2+yJ^2_2),
\eeql{fourfermin1}
in the notation of \cite{Durkin:1985ev,Langacker:1991pg},
where
\beq \begin{split}
\rho_{eff}&=\rho_1 \cos^2 \theta + \rho_2 \sin^2 \theta  \\
w&=\frac{g_2}{g_1}\cos\theta\sin\theta (\rho_1-\rho_2)  \\
y &=\left(  \frac{g_2}{g_1} \right)^2 (\rho_1 \sin^2 \theta + \rho_2 \cos^2 \theta),
\end{split} \eeql{rhoeffwy}
with the mixing angle $\theta$ defined in {Eq.~\ref{rotation}}. For small $\rho_2$ and
$\theta$, these are approximated by
\beq \rho_{eff}\sim \rho_1, \qquad w\sim \widehat \theta, \qquad y \sim \widehat \rho_2,
\eeql{smallrt}
where
\beq
 \widehat \theta\equiv  \frac{g_2}{g_1}  \theta=C \widehat \rho_2, \qquad 
 \widehat \rho_2\equiv \left(  \frac{g_2}{g_1} \right)^2 \rho_2.
\eeql{hatdef}
$C$ is the Higgs-dependent mixing parameter of ${\cal O}(1)$ defined in {Eq.~\ref{mixing2}}.
In the same limit, from {Eq.~\ref{mixing1}},
\beq
\rho_1 \sim \rho_0 (1+\rho_o \theta^2/\rho_2) \xrightarrow[\rho_0=1]{} 1+\theta^2/\rho_2=1+C^2\widehat \rho_2,
\eeql{rho1lim}
 where $\rho_0$, defined in {Eq.~\ref{rhopar}}, is 1 if there are
only Higgs singlets and doublets.

At the $Z$ pole, in addition to the shift in $M_1$ below the SM value, any mixing will
affect the current $\sum_\beta g_\beta  U_{1 \beta}J^\mu_\beta /g_1$ that couples to the $Z_1$.
For $n=1$, the
vector and axial couplings $V_i$ and $A_i$ of the $Z_1$ to fermion $f_i$, which determine the various $Z$ pole asymmetries and
partial widths~\cite{Yao:2006px},
become
\beq \begin{split}
V_i &=\cos \theta  g^1_V(i) + \frac{g_2}{g_1} \sin \theta   g^2_V(i) \sim
g^1_V(i) + \widehat \theta   g^2_V(i)  \\
A_i &=\cos \theta  g^1_A(i) + \frac{g_2}{g_1} \sin \theta   g^2_A(i) \sim
g^1_A(i) + \widehat \theta   g^2_A(i),
\end{split} \eeql{shiftedva}
where $g^\alpha_{V,A}(i)$ are defined in {Eq.~\ref{currents}}.
It should be noted that the $S$, $T$, $U$ formalism~\cite{Peskin:1990zt} only describes
propagator corrections and is not appropriate
for  most  \zpr~s.

\subsubsection{Radiative Corrections}\label{sec:renorm}
The expressions for the electroweak couplings in  {Eq.~\ref{currents}}, {\ref{rhoeffwy}}, and
{~\ref{shiftedva}}
and for $M_{Z_0}$ in {Eq.~\ref{rhoMZ}} are valid at tree level only. One must also
apply full radiative corrections. In practice, since one is searching for very small
tree-level effects from the \zpr\, it is a reasonable approximation to use the SM radiative
corrections~\cite{Yao:2006px} and neglect the effects of the \zpr\ in loops\footnote{The largest
effects are from $Z_2$ loops in $\mu$ decay, which modifiy slightly the relation between the extracted
Fermi constant and the $W$ and $Z$ masses~\cite{Degrassi:1989mu}. $Z_2$ loops
can also modify the relation between $\mu$ and $\beta$ decay and therefore affect the
CKM universality tests~\cite{Marciano:1987ja}.
However, these effects are small for the currently allowed masses.}. However, some care is necessary in the definitions of the renormalized
parameters, e.g., by using the \msb\ rather than the on-shell definition of
\sinn, to ensure that they are not significantly affected by \zpr\ effects~\cite{Degrassi:1989mu,Chankowski:2006jk}.

\subsubsection{Results}\label{sec:results}

\begin{figure*}[htbp]
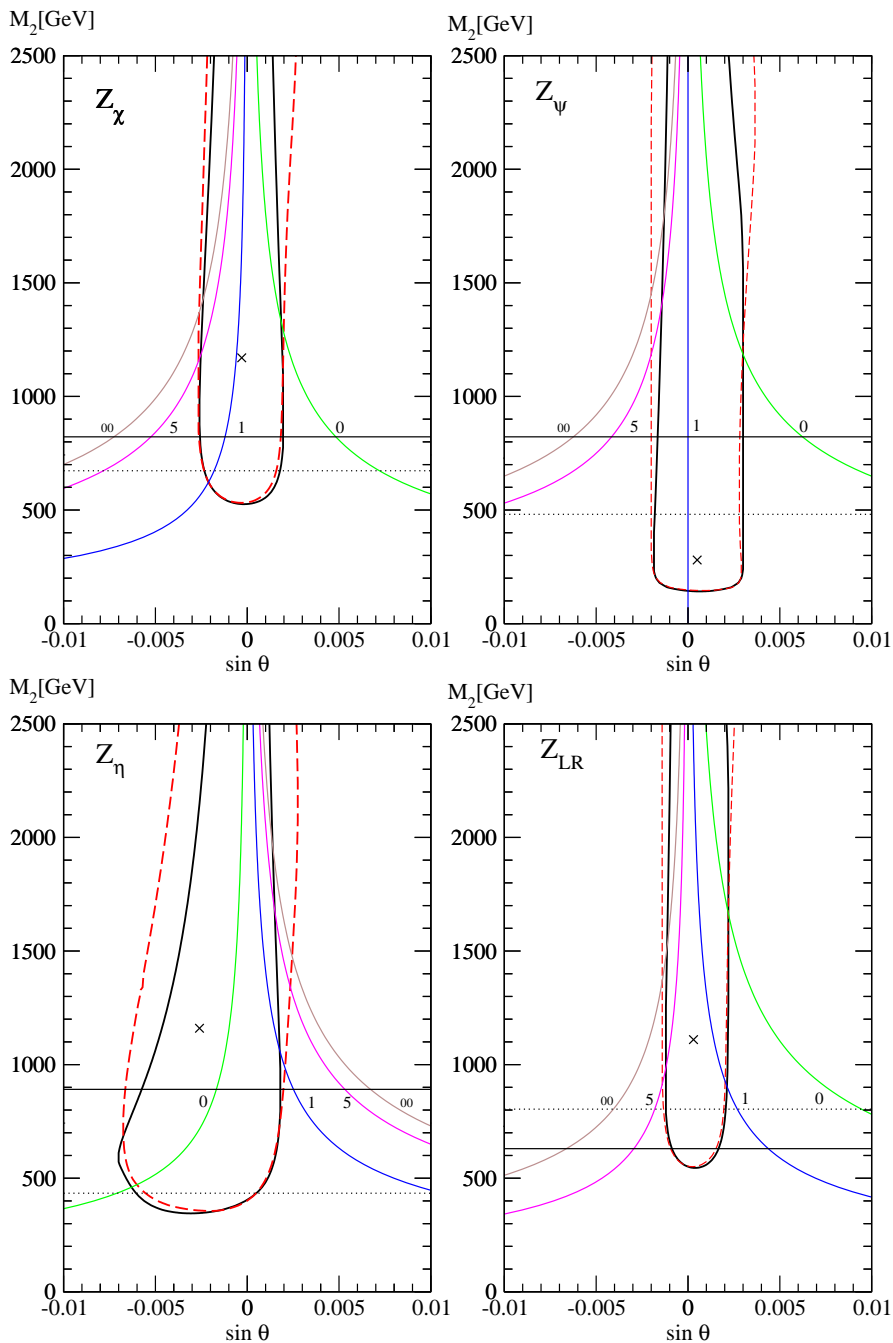

\begin{center}
\includegraphics*[scale=0.5]{chipsi.eps}
\includegraphics*[scale=0.5]{etalr.eps}
\caption{Limits on the \zpr\ mass $M_2$ and the $Z-\zpr$ mixing angle $\theta$
for the $\chi$, $\psi$, $\eta$, and $LR\ (\alpha=1.53)$ models. The solid (dashed) contours
are 90\% cl exclusions from precision electroweak data for $\rho_0=1\ (\rho_0 ={\rm\ free})$.
A cross, $x$, is the best fit. The horizontal solid line is the 95\% cl \ Tevatron lower limit, assuming decays into SM particles only.
The horizontal dotted line is the 95\% cl lower limit from LEP 2. The contours marked $0, 1, 5, \infty$
are for various theoretical relations between the mass and mixing and
are defined in the text. Updated from~\cite{Erler:1999ub}.}
\label{fig:limits}
\end{center}
\end{figure*}

The results from precision electroweak and other data
are shown for some specific models in Table~\ref{tab:limits} and Figure~\ref{fig:limits}.
One sees that the precision data strongly constrain the $Z-\zpr$ mixing angle $\theta$.
They also give lower limits on $M_2$, but these are weaker than the Tevatron and LEP 2 limits.
The precision limit on the $Z_\psi$ mass is low  due to its weak coupling to the neutrino and its
purely axial coupling to the $e^-$.
There is no significant indication for a \zpr\ in the data (although the NuTeV anomaly could possibly be explained by a \zpr\ coupling to $B-3L$~\cite{Davidson:2001ji}). The precision results are presented
for two cases: $\rho_0$ free is for an arbitrary Higgs structure, while $\rho_0=1$ is for Higgs doublets
and singlets with unrestricted charges (i.e., $C$ is left free). There is little
difference between the limits obtained. The precision electroweak constraints are for the
$g_2$ value in {Eq.~\ref{standardcoupling}} (except for the sequential model, which uses
$g_2=g_1\sim 0.74$); for other values the limits on $\theta$ and $M_2$
scale as $g_2^{-1}$ and $g_2$, respectively.

The stringent mixing limits from (mainly) the $Z$ pole data lead to strong indirect limits
on the \zpr\ mass for specific theoretical values of $C$, as can be seen from the theoretical
curves  labeled  $0, 1, 5, \infty$  in Figure  \ref{fig:limits}~\cite{Langacker:1991pg}. For the $\chi$ and LR models the label
refers to the value of $|x|^2/(|\nu_u|^2+|\nu_d|^2)$, where $x/\sqrt 2$ is the VEV of
an extra Higgs doublet that is sometimes considered (transforming like an $L$ doublet
for $\chi$ or like the $\delta^0_L$ defined in Section~\ref{lrmodels} for the LR).
The most commonly studied cases are for $x=0$, which  yield $M_{Z_\chi} > 1368 {\rm\ GeV}, M_{Z_{LR}} > 1673  {\rm\ GeV}$ at 95\% cl.
For the $\psi$ and $\eta$ models, the label represents $\tan^2 \beta \equiv |\nu_u|^2/|\nu_d|^2$,
with $x=0$ assumed.

\begin{table*}[ht]
\caption{95\% cl lower limits on various extra \zpr\ gauge boson masses (GeV) and 90\% cl ranges for
the mixing $\sin \theta$ from precision electroweak data (columns 2-4), Tevatron searches
(assuming decays into SM particles only), and LEP 2.
From~\cite{Erler:1999ub,Alcaraz:2006mx,:2007sb,Yao:2006px}.
The Tevatron numbers in parentheses are preliminary CDF results from March, 2008
based on 2.5 fb$^{-1}$
(CDF note CDF/PUB/EXOTIC/PUBLIC/9160).
 \label{tab:limits}}
\begin{center}
\begin{tabular}{|c|c| c| c| c| c|}
\hline &$\rho_0$ free &$\rho_0=1 $& $\sin \theta\ (\rho_0=1)$ & Tevatron & LEP 2\\
\hline
$\chi$ & 551  & 545 & $(-0.0020)-(+0.0015)$& 822 (864)& 673 \\
$\psi$ & 151  & 146 & $(-0.0013)-(+0.0024)$& 822 (853)& 481 \\
$\eta$ & 379  & 365 & $(-0.0062)-(+0.0011)$& 891 (933)& 434 \\
$LR$ & 570 & 564 & $(-0.0009)-(+0.0017)$& 630 & 804 \\
sequential & 822  & 809 & $(-0.0041)-(+0.0003)$& 923 (966)& 1787 \\
\hline
\end{tabular}
\end{center}
\end{table*}

\subsection{Constraints from Colliders}\label{sec:colliders}
\subsubsection{Hadron Colliders}\label{hadroncollider}
The primary discovery mode for a \zpr\ at a hadron collider is the Drell-Yan
production of a dilepton resonance $p p (\bar p p) \ra \zpr\ra\ell^+\ell^-$,
where $\ell=e$ or $\mu$~\cite{Langacker:1984dc,Barger:1986hd,delAguila:1989be,Leike:1998wr,Dittmar:1996my,Godfrey:2002tna,Dittmar:2003ir,Carena:2004xs,Kang:2004bz,Weiglein:2004hn,:2007sb,Yao:2006px}.
Other channels, such as $\zpr\ra jj$ where $j=$ jet~\cite{Weiglein:2004hn}, $\bar t t$~\cite{Han:2004zh}, $e\mu$~\cite{Abulencia:2006xm}, or $\tau^+\tau^- $,
are also possible. The forward-backward asymmetry for  $p p (\bar p p) \ra\ell^+\ell^-$
(as a function of rapidity, $y$, for $pp$) due to $\gamma,Z,\zpr$ interference below the \zpr\
peak is also important~\cite{Langacker:1984dc,Rosner:1995ft,Dittmar:1996my,:2007sb}.

The cross section for hadrons $A$ and $B$ at CM energy $\sqrt{s}$  to produce a $Z_\alpha$ of mass $M_\alpha$
at rapidity $y$ is, in the narrow width approximation~\cite{Langacker:1984dc},
\begin{multline} \label{hadproduction}
\frac{d\sigma}{dy}= 
 \frac{4\pi^2 x_1x_2}{3M_\alpha^3} \\ \x \sum_{i} (f_{q_i}^A(x_1)f_{\bar q_i}^B(x_2)
+f_{\bar q_i}^A(x_1)f_{q_i}^B(x_2))
\Gamma (Z_\alpha \ra q_i \bar q_i),
\end{multline}
where $f_{q_i,\bar q_i}^{A,B}$ are the structure functions of quark (or antiquark) $q_i$ 
($\bar q_i$) in
hadrons $A$ or $B$, 
and the momentum fractions are
\beq
x_{1,2}=(M_\alpha/\sqrt{s}) e^{\pm y}.
\eeql{xival}
Neglecting mixing effects  the decay width into fermion $f_i$ is
\beq
\Gamma^\alpha_{f_i}\equiv \Gamma (Z_\alpha \ra f_i \bar f_i)= \frac{g_\alpha^2 C_{f_i} M_\alpha}{24\pi} 
\left( \epsilon_L^{{\alpha}}(i)^2+\epsilon_R^{{\alpha}}(i)^2 \right),
\eeql{widthfermion}
where the fermion mass has been neglected. $C_{f_i}$ is the color factor
(1 for color singlets, 3 for triplets). Formulas including fermion mass effects, decays into bosons, 
Majorana fermions, etc.,
are given in~\cite{Kang:2004bz}.

To a good first approximation, {Eq.~\ref{hadproduction}} leads to the \zpr\
total production cross section~\cite{Leike:1998wr}
\beq
\sigma_{\zpr}= {\frac{1}{s}} c_{\zpr} CK \exp(-A{\frac{M_{Z'}}{ \sqrt{s}}}),
\eeql{cros}
where $C$=600 (300) and $A$=32 (20) for $pp$ ($p\bar{p}$)
collisions, and $K \sim 1.3$ is  from higher order corrections.
 From (\ref{cros}),  the predicted cross
section falls exponentially as a function of $M_{Z'}$. The details
of the $Z'$ model are collected in  $c_{\zpr}$, which depends
on $M_{Z'}$, the $Z'$ couplings, and the masses of the
decay products,
\beq
c_{Z'} \equiv  {\frac{4 \pi^2}{ 3}} {\frac{\Gamma_{Z'} }{ M_{Z'}}}
\left( B_u+{\frac{1}{ C_{ud}}} B_d \right), 
\eeql{cz}
where $C_{ud}=2\ (25)$, $\Gamma_{Z'}$ is the total $Z'$ width, and
$B_f=\Gamma_f/\Gamma_{Z'}$ is the branching ratio into $f\bar{f}$.
It is also useful to define 
\beq
\sigma_{\zpr}^f \equiv \sigma_{\zpr} B_f =N_f/\cal L,
\eeql{crosf}
where $N_f$ is the number of produced $f\bar f$ pairs for integrated luminosity $\cal L$.
More detailed estimates for the Tevatron and LHC are given in~\cite{Leike:1998wr,Godfrey:2002tna,Dittmar:2003ir,Carena:2004xs,Fuks:2007gk},
including discussions of parton distribution functions, higher order effects, width effects, 
resolutions, and backgrounds.

The production cross sections, widths, and branching ratios are considered in detail in~\cite{Langacker:1984dc,Barger:1986hd,Kang:2004bz,Gherghetta:1996yr}. For the $E_6$ models, the total width is close to
0.01~\mzp\ assuming decays into SM fermions only and $g_2\sim\sqrt{5/3} g \tan \theta_W$.
However, $\Gamma_{\zpr}$ would be larger if superpartners and/or exotics are light enough
to be produced in the \zpr\ decays, and could therefore be as large as 0.05 \mzp\ in the $E_6$ models~\cite{Kang:2004bz}. The rates for a given channel, such as $\sigma_{\zpr}^e$, decrease as $\Gamma_{\zpr}^{-1}$ in that case. On the other hand, for smaller $\Gamma_{\zpr}$ but fixed branching ratios (e.g., from
some of the decoupled models described in Section~\ref{sec:reduced}) the leptonic rate would decrease
and the peak could be smeared out by detector resolution effects.

\begin{figure*}[htbp]
\begin{center}
\includegraphics*[scale=0.83]{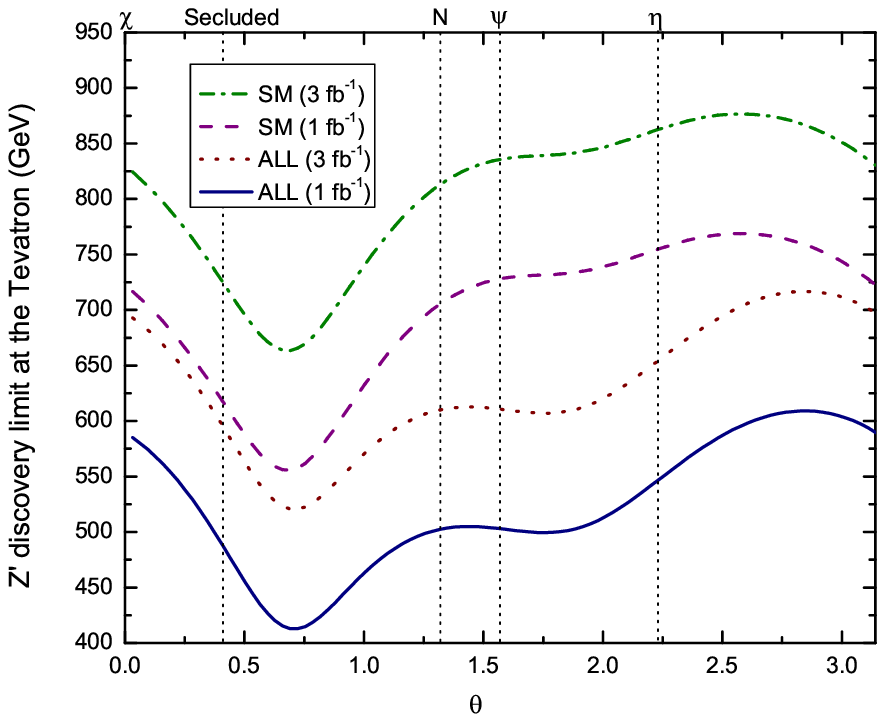}
\includegraphics*[scale=0.83]{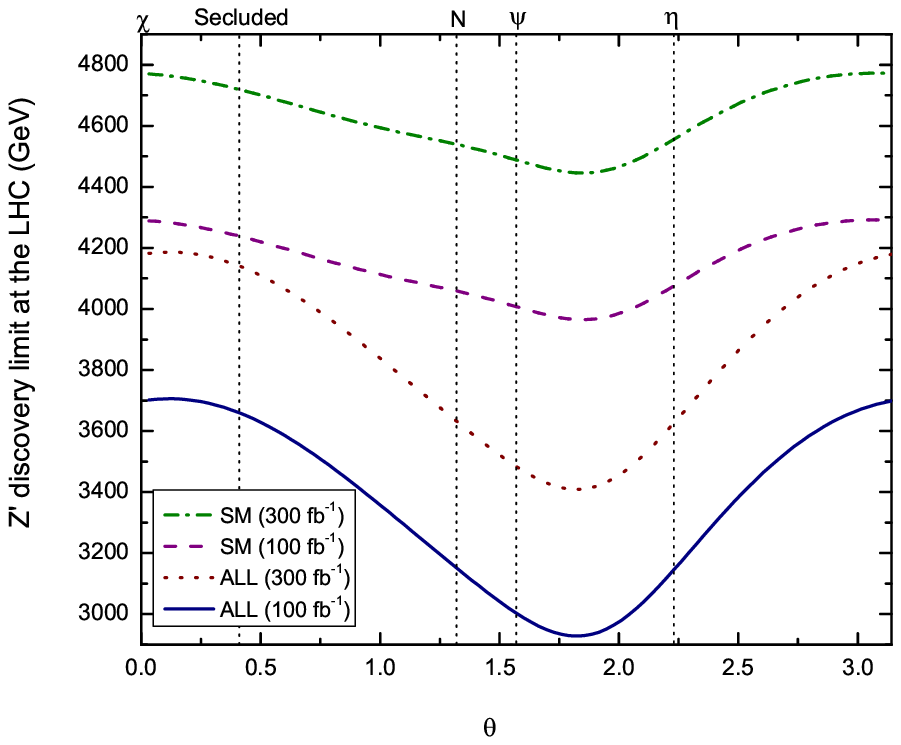}
\caption{Discovery limits for an $E_6$ \zpr\ as a function of  $\theta\equiv\theta_{E_6}$ corresponding to a total of 10 $e^+e^-$ or $\mu^+\mu^-$
events using $\sigma_{\zpr}$ from {Eq.~\ref{cros}}. In each panel the top two
curves assume decays into SM fermions only, while the bottom two assume
that decays into exotics and sparticles are unsuppressed. The different shapes of the
Tevatron and LHC curves is because the $u$ quark dominates at the Tevatron, while
the $u$ and $d$ are more comparable at the LHC.
 From~\cite{Kang:2004bz}.}
\label{fig:hadronreach}
\end{center}
\end{figure*}

The Tevatron limits from the CDF and D0 collaborations~\cite{:2007sb,Yao:2006px}
(dominated, at the time of this writing, by the CDF $e^+e^-$ search  using 1.3 fb$^{-1}$ of data) are given
in Table~\ref{tab:limits}. Figure~\ref{fig:hadronreach} shows  the
sensitivity of the Tevatron and LHC to the $E_6$ bosons as a function of $\theta_{E_6}$
for $\cal L =$ 1 or  3 fb$^{-1}$ (Tevatron), and 100 or 300  fb$^{-1}$ (LHC),
requiring  10 events in the combined $e^+e^-$ and $\mu^+ \mu^-$ channels. 
The Tevatron sensitivity is in the 600-900 GeV range for decays into standard model fermions only, but lower by as much as 200 GeV in the (extreme) case of unsuppressed decays into 
sparticles and exotics. The LHC sensitivity is around 4-5 TeV, but can be lower by $\sim$ 1 TeV
if the sparticle/exotic channels are open.

\subsubsection{$e^+e^-$ Colliders}\label{epepcollider}
\zpr~s much heavier than the CM energy in $e^+e^-$ collisions above the $Z$ pole would manifest
themselves as  new four-fermion interactions analogous to {Eq.~\ref{fourfermi},
but with the $\alpha$ sum starting at 2. These would interfere with the virtual
$\gamma$ and $Z$ contributions for leptonic and hadronic final states (see, e.g.,~\cite{Cheung:2001wx}).

The ALEPH, DELPHI, L3, and OPAL collaborations at LEP2 have measured
production cross sections and angular distributions or asymmetries for $e^+e^-\ra e^+e^-, \mu^+\mu^-, 
\tau^+ \tau^-, \bar c c$, and $\bar b b$, as well as hadronic cross sections,  at CM energies up to $\sim$ 209 GeV~\cite{Alcaraz:2006mx}. 
They saw no indication of new four-fermi interactions,  and the combined lower limits
for typical models are given in Table~\ref{tab:limits} and Figure~\ref{fig:limits}.

Similarly, a future linear collider  would have sensitivity to \mzp\ 
well above the CM energy by interference with the $\gamma$ and $Z$~\cite{delAguila:1993rw,Leike:1996pj,Babich:1998ri,Richard:2003vc,Godfrey:2005pm,Weiglein:2004hn}.
Observables  could include production cross sections, forward-backward (FB) asymmetries,
polarization (LR) asymmetries, and mixed FB-LR asymmetries
for $e^+e^-\ra e^+e^-, \mu^+\mu^-, \tau^+\tau^-, \bar c c, \bar bb$, and $\bar t t$;
 $\tau$ polarization; and  cross sections and polarization asymmetries for $\bar q q$.
 High luminosity, $e^-$ polarization, and efficient tagging of heavy flavors are important.
 For example, the International Linear Collider (ILC) with $\sqrt{s}=500$ GeV,
 ${\cal L} = 1000 {\rm \ fb}^{-1}$, and $P_{e^-}=80$\% would have 5$\sigma$ sensitivity to
 the $E_6$ and $LR$ bosons in the range $2-5$ TeV, increasing by $\sim 1$ TeV for
  $\sqrt{s}=1$ TeV~\cite{Weiglein:2004hn}.  There is some chance that a \zpr\ could be observed first
  at the ILC, e.g., if its mass were beyond the LHC range or its couplings weak,
  in which case only  $M_2/g_2$ could be determined  for large $M_2$.
  More likely, the \zpr\ would be discovered first and \mzp\ determined independently at  the LHC
  or Tevatron.  A GigaZ ($Z$-pole) option for the ILC would be extremely sensitive to
  $Z-\zpr$ mixing~\cite{Weiglein:2004hn}.

\subsection{Diagnostics of \zpr\ Couplings}\label{sec:diagnostics}
Following the discovery of a resonance in the $\ell^+\ell^-$ channels, the next step would be
to establish its spin-1 nature (as opposed, e.g., to a spin-0 Higgs resonance or a spin-2
Kaluza-Klein graviton excitation). This can be done by the angular distribution in the 
resonance rest frame, which for spin-1 is
\beq
\frac{d \sigma_{\zpr}^f }{d \cos \theta^\ast}  \propto \frac{3}{8} (1+\cos^2\theta^\ast)+A_{FB}^f \cos \theta^\ast,
\eeql{angdist}
where $\theta^\ast$ is the angle between the incident quark or lepton and fermion $f$.
Of course, for a hadron collider one does not know which hadron is the source of the $q$
and which the $\bar q$ on an event by event basis, but 
the ambiguity washes out in the determination
of the $1+\cos^2\theta^\ast$ distribution characteristic of spin 1~\cite{Langacker:1984dc,Dittmar:1996my}. The spin can also be probed in $e^+e^-$ by polarization asymmetries~\cite{Weiglein:2004hn}.

One would next want to determine the chiral couplings to the quarks, leptons, and other particles in order
to discriminate between models. (The gauge coupling $g_2$ can be fixed to the value in
{Eq.~\ref{standardcoupling}}, or alternatively can  be taken as a free parameter if the charges are normalized by some convention.)
This should be possible for masses up to $\sim$ $2-2.5$ TeV at the LHC assuming  typical couplings, but for higher masses there are
too few events for meaningful diagnostics. 

 In the main LHC production channels,  $pp \to Z'\to \ell^+
\ell^-$ ($\ell=e,\mu$), one would be able to measure
the mass $M_{Z'}$, the width $\Gamma_{\zpr}$ and the leptonic cross section
 $\sigma_{\zpr}^{ \ell}= \sigma_{\zpr} B_\ell$. By itself, $\sigma_{\zpr}^{ \ell}$  is { not}   a useful diagnostic
for the $Z'$ couplings to quarks and leptons:
while $\sigma_{\zpr}$   can be
calculated to within  a few percent for
given $Z'$ couplings, the branching ratio into leptons,
$B_\ell$,
depends strongly on the contribution of exotics
 and sparticles to $\Gamma_{\zpr}$~\cite{Kang:2004bz}.
However, $\sigma_{\zpr}^{ \ell}$
would be a useful indirect probe for the existence of the exotics
 or superpartners.
Furthermore, the product
$\sigma_{\zpr}^{ \ell}\Gamma_{\zpr} = 
\sigma_{\zpr} \Gamma_\ell$ does
 probe the absolute magnitude of the quark and lepton couplings.

The most useful diagnostics   involve the {relative  
strengths} of $Z'$
couplings to ordinary quarks and leptons. The  
forward-backward asymmetry as a function of the $Z'$
rapidity, $A_{FB}^f(y)$~\cite{Langacker:1984dc},
avoids the $\bar q q$ ambiguity in {Eq.~\ref{angdist}}.
For $AB\ra\zpr\ra \bar f f$, define $\theta_{CM}$ as the angle of fermion $f$
with respect to the direction of hadron $A$
in the \zpr\ rest frame, and let $F$ ($B$) be the cross section for fixed rapidity $y$
with $\cos \theta_{CM} >0$ ($<0$). Then, $A_{FB}^f(y)\equiv(F-B)/(F+B)$, with
\beq \begin{split}
F\pm B&\sim \left[\begin{array}{c} 4/3 \\ 1\end{array} \right]
 \sum_{i} \left( f_{q_i}^A(x_1)f_{\bar q_i}^B(x_2)\pm f_{\bar q_i}^A(x_1)f_{q_i}^B(x_2) \right) \\
&\x  \left( \epsilon_L(q_i)^2\pm\epsilon_R(q_i)^2 \right)
 \left( \epsilon_L(f)^2\pm\epsilon_R(f)^2 \right).
\end{split} \eeql{abfy} 
 Clearly, $A_{FB}^f(y)$ vanishes for $pp$ at $y=0$,
but can be nonzero at large $y$ where there is more likely a valence $q$ from the first proton
and sea $\bar q$ from the other. The leptonic forward-backward asymmetry is
sensitive to a combination of quark and lepton chiral couplings and is 
a powerful discriminant between models~\cite{Langacker:1984dc}.
 
There are a number of additional  probes. The 
ratio of cross sections in different rapidity bins~\cite{delAguila:1993ym} gives
information on the relative $u$ and $d$ couplings. Possible observables in other two-fermion final state channels
include the polarization of produced $\tau$'s~\cite{Anderson:1992jz}
and the $pp\to Z'\to j  j$ cross section~\cite{Rizzo:1993ie,Weiglein:2004hn}.
There are no current plans for polarization at the LHC, but polarization asymmetries
at a future or upgraded hadron collider would provide another useful diagnostic~\cite{Fiandrino:1992fa}.

 In four-fermion final state  channels the
rare decays $\zpr\ra V f_1 \bar f_2$, where $V=W$ or $Z$ is radiated from the \zpr\ decay products, have a double logarithmic enhancement. 
In particular, $Z'\to W\ell\nu_{\ell}$ (with $W\ra$ hadrons and an  $\ell\nu_{\ell}$ transverse
mass  $> 90$ GeV to separate from SM background) may be observable
and projects out the left-chiral lepton couplings~\cite{Rizzo:1987rw,Cvetic:1991gk,Hewett:1992nf}.
Similarly, the associated productions $pp\rightarrow Z' V$
with $V=(Z,W)$~\cite{Cvetic:1992qv}  and $V=\gamma$~\cite{Rizzo:1992sh}
could yield information on the quark chiral couplings.

Finally, decays into two bosons, such as $\zpr\ra W^+ W^-, Zh,$ or $W^\pm H^\mp$,
can occur only by $Z-\zpr$ mixing or with amplitudes related to the mixing. 
However, this suppression may be compensated for the longitudinal
modes of the $W$ or $Z$ by the large polarization vectors, with components scaling as
$\mzp/M_W$~\cite{delAguila:1986ad,Barger:1987xw,Deshpande:1988py}.
For example, $\Gamma(\zpr \ra W^+ W^-)\sim \theta^2,$ which appears to be hopelessly
small to observe. However, the enhancement factor is $\sim (\mzp/M_W)^4$. Thus, from
{Eq.~\ref{mixing2}}, these factors compensate, leaving a possibly observable rate that
in principle could give information on the Higgs charges. In the limit of $\mzp \gg M_Z$ one has
\beq
\Gamma(\zpr \ra W^+ W^-) = \frac{g_1^2 \theta^2\mzp}{192\pi} \left(\frac{\mzp}{M_Z}\right)^4
= \frac{g_2^2 C^2 \mzp}{192\pi}.
\eeql{zpww}

Global studies of the possible LHC diagnostic possibilities for determining ratios
of chiral charges in a model independent way and discriminating models are given in~\cite{Cvetic:1995zs,delAguila:1993ym}. The complementarity of LHC and ILC observations is
especially emphasized in~\cite{delAguila:1993rw,Cvetic:1995zs,delAguila:1995fa,Weiglein:2004hn}.

\section{Implications}\label{sec:implications}

\subsection{The $\mu$ Problem and Extended Higgs/Neutralino Sectors}\label{sec:muHiggs}
\subsubsection{The $\mu$ Problem}\label{sec:mu}
As described in Section~\ref{muproblem}, the $\mu$ problem of the MSSM can be solved
in {\em singlet extended} models in which a symmetry forbids an elementary $\mu$ term, but allows
a dynamical $\mu_{eff}=\lambda_S \langle S \rangle$.
There are a number of realizations of this mechanism  (see~\cite{Accomando:2006ga,Barger:2007ay} for reviews). The best known is the {\em next to minimal model} (NMSSM), in which 
 a discrete $Z_3$ symmetry forbids $\mu$ but allows the cubic terms $ \lambda_S S H_u H_d$
and $\kappa S^3/3$ in the superpotential~\cite{Ellis:1988er}. The original form of the NMSSM
suffers from cosmological domain wall problems because of the discrete symmetry. This can
be remedied in more sophisticated forms involving an R symmetry~\cite{Accomando:2006ga}. A variation on that
approach yields the {\em new minimal model} (nMSSM), in which the cubic term
and its soft analog are replaced by tadpole terms linear in $S$ with sufficiently small
coefficients~\cite{Panagiotakopoulos:1999ah}. A \uprm\ symmetry, which  is perhaps more likely to emerge from a string construction,
is another possibility~\cite{Cvetic:1995rj,Suematsu:1994qm,Cvetic:1997ky}. This avoids the domain wall problem by embedding the
discrete symmetry of the NMSSM into a continuous one.

\subsubsection{Extended Higgs Sector}\label{extendedHiggs}

Conventional \uprm\ models necessarily involve extended Higgs sectors associated with
the SM singlet fields whose VEVs break the \uprm\  symmetry.  Especially interesting in this respect are
those supersymmetric models involving 
a dynamical $\mu_{eff}=\lambda_S \langle S \rangle$.
If one ignores Higgs sector CP violation\footnote{Loop effects may generate
significant CP effects, especially for the heavier Higgs states~\cite{Demir:2003ke}.}, then there will be an additional Higgs scalar associated with
 $S$, that can mix with the two MSSM scalars from $H^0_{u,d}$.
 (There is
also an additional pseudoscalar in the models involving a discrete symmetry.)
  Since the $S$ does not
 couple directly to the SM fermions or gauge bosons, the LEP {\em lower} limits on the
 Higgs mass ($m_H > 114.4$ GeV for the SM Higgs, and somewhat weaker in the MSSM)
 are weakened if the lightest Higgs has a significant singlet component~\cite{Barger:2006dh}.
 Conversely, the theoretical {\em upper} limit on the lightest Higgs is also
 relaxed, from $\sim 130$ GeV in the MSSM to around $170$ GeV in the simplest \uprm\ model,
 due to the new $F$ and $D$ term contributions to the potential in {Eq.~\ref{MSSMUV}}.
 (One must include the loop corrections to these estimates~\cite{Barger:2006dh}.) 
 These relaxed limits   allow lower values for $\tan \beta \equiv \nu_u/\nu_d$
 in the \uprm\ models than are favored for the MSSM.
 
The {\em UMSSM} is the \uprm\ with a single $S$, with the potential in {Eq.~\ref{MSSMUV}}.
In the decoupling limit, $\langle S \rangle \ra \infty$ with $\mu_{eff}$ fixed, the UMSSM reduces to the
MSSM. Existing constraints
favor this limit (unless the \zpr\ is leptophobic, with small $Z-\zpr$ mixing
due to a cancellation of the two terms in $\Delta^2$ in {Eq.~\ref{examplepars}}).
For large $\langle S \rangle$ the extra Higgs is heavy and mainly singlet~\cite{Barger:2006dh},
so the Higgs sector is MSSM-like.
However, more general \uprm\ models such as the secluded model
in Section~\ref{sec:intermediate}, as well as other models such as the nMSSM,
can yield significant doublet-singlet mixing, light singlet-dominated states, etc~\cite{Erler:2002pr,Han:2004yd}.
(In fact, the secluded model reduces to the nMSSM in an appropriate limit~\cite{Barger:2006dh}.)
This may yield such nonstandard collider signatures as light weakly coupled Higgs,
multiple Higgs with reduced couplings, and invisible decays into light neutralinos~\cite{Barger:2006sk}.

\subsubsection{Extended Neutralino Sector}\label{sec:neutralino} 
The neutralino sector of the MSSM (the bino, $\tilde B$, and the wino, $\tilde W^0$, with soft masses
$M_{\tilde B,\tilde W}$; and two neutral Higgsinos $\tilde H_{u,d}^0$) is extended in
\uprm\ models by one or more singlinos, $\tilde S$, and by the \zpr-gaugino, $\tilde Z'$,
with soft mass $M_{\tilde Z'}$~\cite{Suematsu:1997au,Hesselbach:2001ri,Barger:2005hb,Barger:2006kt,Choi:2006fz}. (There could also be soft mass or kinetic $\tilde B-\tilde Z'$ mixing terms.)
  
  In the $(\tilde B,\tilde W^0,\tilde H_d^0,\tilde H_u^0, \tilde S,\tilde Z')$ basis, the mass matrix 
  for the six neutralinos in the UMSSM is
  \begin{widetext}
\beq
{M_{\chi^{0}}}=
\left( \begin{array}{cccccc}
M_{\tilde B} & 0 & -g'\nu_d/2 & g'\nu_u/2 & 0 & 0 \\
0 & M_{\tilde W} & g\nu_d/2 & -g\nu_u/2 & 0 & 0 \\
-g'\nu_d/2 & g\nu_d/2 & 0 & -\mu_{eff} & 
-\mu_{eff}\nu_u/s & g_2Q_{d}\nu_d \\
g'\nu_u/2 & -g\nu_u/2 & -\mu_{eff} & 0 & 
-\mu_{eff}\nu_d/s & g_2Q_{u}\nu_u \\
0 & 0 & -\mu_{eff}\nu_u/s & -\mu_{eff}\nu_d/s & 
0 & g_2Q_{S}s \\
0 & 0 & g_2Q_{d}\nu_d & g_2Q_{u}\nu_u & 
g_2Q_{S}s & M_{\tilde Z'} \\
\end{array} \right).
\eeql{eqn:neutmtx}
  \end{widetext}
In the decoupling limit with  $g_2Q_{S}s \gg M_{\tilde Z'}$
the singlino and  the \zpr-gaugino will combine to form an approximately
Dirac fermion with mass $g_2Q_{S}s \sim \mzp$ and little mixing with the four MSSM 
neutralinos. For large $ M_{\tilde Z'}\gg g_2Q_{S}s$, on the other hand, there
will be a heavy Majorana $\tilde Z'$, and a much lighter singlino with a seesaw
type mass $\sim -\mzp^2/ M_{\tilde Z'}$. For smaller $s$ there can be significant
mixing with the MSSM neutralinos. One can easily extend to secluded models~\cite{Erler:2002pr,Han:2004yd}, models with multiple \uprm~s~\cite{Hesselbach:2001ri},
or singlet extended models with discrete symmetries~\cite{Accomando:2006ga,Barger:2007ay,Barger:2005hb}.
In many of these cases there are light singlino-dominated states (which can be the
lightest supersymmetric particle (LSP)) and/or significant
mixing effects. These can lead to a variety of collider signatures very different from the
MSSM~\cite{Barger:2006kt}. For example, in some cases there are four MSSM-like
neutralinos with production and cascades similar to the MSSM. However, the lightest
of these may then undergo an additional decay to a singlino LSP, accompanied e.g., by
an on-shell $Z$ or Higgs. Enhanced rates for the decay of chargino-neutralino pairs to
three or more leptons are possible. It is also possible for the lightest Higgs to decay
invisibly to two light singlinos. Cold dark matter implications are described in Section
\ref{sec:cosmology}.

\subsection{Exotics}\label{sec:exotics}

Almost all \uprm\ models require the addition of new chiral exotic states to
cancel anomalies (Section~\ref{sec:anomalies}). Precision electroweak  
constraints favor that these are {\em quasi-chiral}, i.e., vector pairs under the SM but 
chiral under \uprm. Examples are the $SU(2)$-singlet $D, D^c$ 
quarks with  charge $-1/3$  in the $E_6$ model (Table~\ref{tab:e6decomp}); the $SU(2)$ doublet pairs
in $E_6$ which may be interpreted either as additional Higgs pairs $H_{u.d}$ or
as exotic lepton doublets; or SM singlets. Realistic models must provide means of generating
masses for such exotics, e.g., by coupling to chiral (or nonchiral) singlets which
acquire VEVs, such as $S D D^c$,  and also for their decays.

  Consider the example of the exotic $D$ quarks, which can be pair-produced
by QCD processes at a hadron collider, and their scalar supersymmetric partners
$\widetilde D$, produced with an order of magnitude smaller cross section.
(The rates are smaller for exotic leptons.) Once produced, there
are three major decay possibilities~\cite{Kang:2007ib}:
\begin{itemize}
\item The decay may be $D\ra u_iW^-, \ D \ra d_i Z$, or $ D \ra d_i H^0$, if driven by  mixing
with a light charge $-1/3$ quark~\cite{Barger:1985nq,Andre:2003wc}. The current limit is  $m_D \gtrsim$ 200 GeV~\cite{Andre:2003wc},
which should be improved to $\sim$ 1 TeV at the LHC. However, 
such mixing is forbidden in the supersymmetric $E_6$ model if $R$-parity is 
conserved.
\item One may have $\widetilde D\ra j j$ if there is a small diquark operator such as $ u^c d^c D^c$, 
or $\widetilde D \ra \ j\ell$ for a leptoquark operator like $ L Q  D^c$.
(They cannot both be present because of proton decay.)
Such operators do not by themselves violate $R$ parity ($R=+1$ for the scalar), 
and therefore allow a stable lightest
supersymmetric particle. They are strongly constrained by the $K_L-K_S$ mass difference
and by $\mu-e$ conversion, but may still be significant~\cite{Kang:2007ib}.
If the scalar $\widetilde D$ is heavier than the fermion, then it may decay resonantly into
the fermion pair, or into a $D$ and neutralino (or gluino). The lighter fermion $D$ can
decay into a neutralino and nonresonant fermion pair via a virtual $\widetilde D$ or via
a real or virtual  squark or slepton.
A heavier fermion will usually decay into an on-shell $\widetilde D$ and a neutralino (or gluino),
with  the $\widetilde D$ decaying to fermions. The signals from these decays, especially for a heavier
scalar, may be difficult to extract from normal SUSY cascades, especially for diquarks.
However, there are some possibilities
based on missing transverse energy, lepton multiplicities and $p_T$, etc~\cite{Kang:2007ib}.
\item They may be stable at the renormalizable level due to the \uprm, or to an accidental 
or other symmetry,
so that they hadronize and escape from or stop in the detector~\cite{Kang:2007ib},
with signatures~\cite{Kraan:2005ji} somewhat similar to the quasi-stable gluino
expected in split supersymmetry~\cite{ArkaniHamed:2004fb}. They could then decay
by higher-dimensional operators on a time scale of $\lesssim 10^{-1}-100$ s, short enough to
avoid cosmological problems~\cite{Kawasaki:2004qu}. These operators could allow direct
decays to SM particles, or they could involve SM singlets with VEVs which could induce
tiny mixings with ordinary quarks.
\end{itemize}

Exotics carrying SM charges significantly modify the running of the SM couplings, and therefore
can affect gauge unification unless, e.g.,  they occur in $SU(5)$-type multiplets. Examples of \uprm\ constructions which preserve the MSSM
running at tree level are described in Sections \ref{sec:e6model} and \ref{sec:anomfree}.

\subsection{The \zpr\ as a Factory}\label{sec:factory}
The decays of a  \zpr\ could serve as an efficient source of other particles if it is sufficiently massive. This has been explored in
detail for slepton production, $pp\ra\zpr\ra \widetilde \ell \widetilde \ell^\ast$, with
$\widetilde \ell\ra \ell+$ LSP, assuming that \mzp\ is already known from the conventional
$\ell^+\ell^-$ channel~\cite{Baumgart:2006pa}. This can greatly extend the discovery reach
of the $ \widetilde \ell$ and may give information on the identity of the LSP. 
Decays of the \zpr\ could also be a useful production mechanism for pairs of exotics~\cite{Rosner:1996eb}
or heavy Majorana neutrinos~\cite{Duncan:1986wz,delAguila:2007ua}. The latter could lead to the interesting signature of like sign leptons $+$ jets. The total width $\Gamma_{\zpr}$,
in combination with other constraints on the quark and lepton charges, would also give
some information on the exotic/sparticle decays~\cite{Gherghetta:1996yr,Kang:2004bz}.

\subsection{Flavor Changing Neutral Currents}\label{sec:fcnc}
In Section~\ref{sec:basic} it was implicitly assumed that the \uprm\ charges were family
universal. That implies that the \zpr\ couplings are unaffected by fermion mixings
and remain diagonal (the GIM mechanism). However, many models involve nonuniversal charges, 
as described in Section~\ref{sec:nonuniversal}. Let us rewrite the \uprm\ current in
{Eq.~\ref{currents}} as
\beq
J^\mu_\alpha= \bar f^0_L \gamma^\mu \epsilon_{fL}^{{\alpha}} f^0_L+ \bar f^0_R \gamma^\mu \epsilon_{fR}^{{\alpha}} f^0_R,
\eeql{nonuniversalcharge}
where $f^0_L$ is a column vector of weak-eigenstate left chiral fermions of a given type (i.e., $u^0_L, d^0_L, e^0_L$, or $\nu^0_L$), and similarly for $f^0_R$. The $\epsilon^\alpha_f$ are
 diagonal  matrices of \uprm\ charges.
The $f^0_{L,R}$ are related to the mass eigenstates $f_{L,R}$ by 
\beq f^0_{L}=V^{f\dagger}_L f_L, \qquad  f^0_{R}=V^{f\dagger}_R f_R, \eeql{massfermions}
where $V^f_{L,R}$ are unitary. In particular, the CKM and PMNS
matrices are given by $V^u_LV^{d\dagger}_L$ and $V^\nu_LV^{e\dagger}_L$,
respectively. In the mass basis,
\beq
J^\mu_\alpha= \bar f_L \gamma^\mu B_{fL}^{{\alpha}} f_L+ \bar f_R \gamma^\mu B_{fR}^{{\alpha}} f_R,
\eeql{nonuniversalmass}
where
\beq
 B_{fL}^{{\alpha}}\equiv V^f_L  \epsilon_{fL}^{{\alpha}} V^{f\dagger}_L, \qquad
 B_{fR}^{{\alpha}}\equiv V^f_R  \epsilon_{fR}^{{\alpha}} V^{f\dagger}_R.
\eeql{Bdef}
For family universal charges, $\epsilon_{fL,R}$ are proportional to the identity, and
$B_{fL,R}=\epsilon_{fL,R}$. However, for the nonuniversal case, $B_{fL,R}$ will in
general be nondiagonal. As a simple two family example, if
$\epsilon= 
\begin{pmatrix}0&0 \\0 & 1\end{pmatrix}$
and $V$ is a rotation of the same form as {Eq.~\ref{rotation}}
then
\beq
\begin{split}
J^\mu&=\sin^2 \theta  \bar f_1 \gamma^\mu f_1+ \cos^2 \theta  \bar f_2 \gamma^\mu f_2\\
&+ 
\sin \theta \cos \theta ( \bar f_1 \gamma^\mu f_2+  \bar f_2 \gamma^\mu f_1).
\end{split}
\eeql{nonuniversalexample}

The formalism for FCNC mediated by \zpr, and also by off-diagonal $Z$ couplings induced
by $Z-\zpr$ mixing, was developed in~\cite{Langacker:2000ju},
and limits were obtained for a number of tree and loop level mixings and decays. The limits
from $K^0-\bar K^0$ mixing (including CP violating effects) and from $\mu-e$ conversion
in muonic atoms are sufficiently strong to exclude
significant nonuniversal effects for the first two families for a TeV-scale \zpr\ with electroweak
couplings. However, nonuniversal couplings for the third family are still possible
and could contribute~\cite{Langacker:2000ju,Leroux:2001fx,Barger:2003hg,Barger:2004qc,Barger:2004hn,Chen:2006vs,He:2006bk,Chiang:2006we,Baek:2006bv,Cheung:2006tm}  
to processes such as $B\bar B$ and $D \bar D$ mixing, $B\ra \mu^+ \mu^-$, or
$b\ra ss\bar s$ (such as in $B\ra \phi K$).
Since the \zpr\ effects are at tree level, they may be important even for small couplings since they
are competing with SM or MSSM loop effects. 
The possible anomaly observed in the $Z\ra \bar b b$ forward-backward asymmetry~\cite{Yao:2006px}
could possibly be a (flavor diagonal) result of a nonuniversal \zpr\ coupling~\cite{Erler:1999nx}.
Collider processes such as single
top production could possibly be observable as well~\cite{Arhrib:2006sg}.

The nonuniversal couplings could also be relevant to loop effects, such as $b\ra s \gamma$
or $\mu\ra e \gamma$,
or intrinsic magnetic or electric dipole moments. One interpretation of  the possible anomaly~\cite{Yao:2006px} suggested by the BNL experiment for the anomalous magnetic moment of the $\mu$
involves the vertex diagram with a \zpr\ exchange (see, e.g.,~\cite{Cheung:2006tm}).
The flavor-diagonal diagram with an internal $\mu$ is too small to be relevant (unless $\mzp\sim 100$ GeV or the couplings are large). However, an internal $\tau$ enhances the effect by $m_\tau/m_\mu$, and the anomaly could be accounted for by a TeV scale \zpr\ with large  $\mu-\tau$ mixing.

Mixing between the ordinary and exotic fermions can also lead to FCNC effects~\cite{Langacker:1988ur}. For example, a small
$d^c-D^c$ mixing in the $E_6$ model of Table~\ref{tab:e6decomp} would induce 
off diagonal couplings of the $Z^0_2$  to the light and heavy mass eigenstate, while a $d-D$ mixing (i.e., between an $SU(2)$ doublet and singlet) would generate similar effects for the ordinary $Z^0$.
Off-diagonal vertices between the light mass eigenstates, such as $Z^0_\alpha \bar b s$, would be
induced as second order effects.

\subsection{Supersymmetry Breaking, \zpr\ Mediation, and the Hidden Sector}\label{sec:hidden}
     
\uprm~s have many possible implications for supersymmetry breaking and mediation, and
for communication with a hidden sector. For example, one limit of the single $S$
scenario of Section~\ref{sec:tevscale} requires large (TeV scale) soft masses
 in the Higgs sector, suggesting the possibility of heavy sparticles as well~\cite{Everett:2000hb}.

Another implication is the \uprm\ $D$ term
contribution to the scalar potential~\cite{Kolda:1995iw}, 
\beq V_D=\oh D^2 \equiv \frac{1}{2} \left( -g_2\sum_i Q_i |\phi_i|^2 \right)^2.
\eeql{dterm}     
$V_D$ of course contributes to the minimization conditions and Higgs sector masses.
Assuming a value $D^{min} \ne 0$ for $D$ at the minimum, it gives a contribution to
the masses $m_i^2$ of the squarks, sleptons, and exotic scalars 
\beq 
\Delta m^2_i = (- D^{min} )(g_2 Q_i).
\eeql{massshift}
For a single $S$ field, one has 
$- D^{min}=g_2(Q_u |\nu_u|^2 + Q_d|\nu_d|^2+Q_S |s|^2)/2$
in the notation of Section~\ref{sec:ssb}. $\Delta m^2_i $ can be of either sign,
and must be added to other supersymmetric and soft contributions.
When the \uprm\ scale is large, there is a danger of overall negative mass-squares
which de-stabilize the vacuum. 
However, in that case there is the possibility of
breaking along a $D$ flat direction in which $V_D^{min}$ is small, as in the secluded models~\cite{Erler:2002pr}. Positive $D$ term contributions to the slepton masses have been suggested as
a means of compensating the negative ones from anomaly mediated supersymmetry
breaking~\cite{Murakami:2003pb,Anoka:2004vf}.
The $D$ term quartic interactions also contribute to the RGE equations for the soft
masses (see, e.g.,~\cite{Cvetic:1997ky,Langacker:1998tc}).

\uprm~s have been invoked in  many models of supersymmetry breaking or mediation.
For example, many models of gauge mediation involve a \uprm\, which may help transmit
the breaking by loop effects  and/or $D$ terms in the hidden or ordinary 
sectors~\cite{Dobrescu:1997qc,Kaplan:1998jk,Cheng:1998nb,Cheng:1998hc,Langacker:1999hs}.
The Fayet-Iliopoulos terms (Section~\ref{sec:anomsection}) associated with anomalous \uprm~s in string constructions
may also help trigger and transmit supersymmetry breaking~\cite{Mohapatra:1996in,Dvali:1997sr}.

In many string constructions particles in both the ordinary and hidden sector
may carry \uprm\ charges, allowing for the possibility of {\em \zpr\ mediation}~\cite{Langacker:2007ac}.
The simplest case is that the \uprm\ gauge symmetry is not broken in the hidden sector,
but the \zpr\ gaugino acquires a mass from the SUSY breaking. The $\zpr-\tilde Z'$ mass difference
induces ordinary sector scalar masses at one loop, and SM gaugino masses at
two loops. Requiring the latter to be in the range $10^2-10^3$ GeV
implies $M_{\tilde Z'}\gtrsim 10^3$ TeV (for electroweak couplings),
with the sparticles, exotics, and \zpr\ around $10-100$ TeV and the electroweak scale
obtained by a fine-tuning, i.e.,  a version of split supersymmetry~\cite{ArkaniHamed:2004fb}.
String embeddings of this scenario are addressed in~\cite{Verlinde:2007qk}. It can
also be combined with other mediation scenarios, allowing a lower \zpr\ scale~\cite{Nakayama:2007je}.
A \zpr\ communicating with a hidden sector could also allow the production and decays
into SM particles of relatively light {\em hidden valley} particles~\cite{Han:2007ae}.

\subsection{Neutrino Mass}\label{sec:neutrino}
The seesaw model (see, e.g.,~\cite{Mohapatra:2005wg})
leads to a small Majorana mass $m_\nu \sim -m_D^2/M_{\nu^c}$ for the ordinary doublet
neutrinos $\nu$, where $m_D$ is a Dirac mass (generated by the VEV of a Higgs doublet),
and $M_{\nu^c}\gg m_D$ is the Majorana mass of the heavy singlet $\nu^c$,
\beq
-L_\nu = m_D \bar \nu_L \nu_R + \oh M_{\nu^c}\bar \nu^c_L \nu_R + h.c.,
\eeql{seesaw}
where $\nu_R$ is the conjugate of $\nu^c_L$. For $m_D\sim 100$ GeV and
$M_{\nu^c} \sim 10^{14}$ GeV one obtains $|m_\nu| $ in the observed 0.1 eV range.
However, if the $\nu^c$ is charged under a \uprm\ then $M_{\nu^c} $ 
cannot be much larger than the \uprm\ scale.
One possibility for a TeV scale \zpr\ is that the $\nu^c$ is neutral, as in the $N$ model~\cite{Ma:1995xk,Barger:2003zh,Kang:2004ix,King:2005jy}. Then, a conventional seesaw~\cite{Ma:1995xk,Keith:1996fv,Kang:2004ix,King:2005jy}
and leptogenesis~\cite{Hambye:2000bn} scenario can be possible
if a large $M_{\nu^c}$ can be generated. For other models
 with TeV scale \mzp\
one must invoke an alternative to the seesaw. For example,
small Majorana masses can be generated using the double seesaw
mechanism (involving an additional power of $M_{\nu^c}^{-1}$), or
by invoking a Higgs triplet~\cite{Kang:2004ix,Mohapatra:2005wg}.

Another possibility, which can lead to either small Dirac or Majorana masses,
involves higher-dimensional operators (HDO)~\cite{Cleaver:1997nj,Langacker:1998ut,ArkaniHamed:2000bq,Borzumati:2000mc,Gogoladze:2001kj,Kang:2004ix,Chen:2006hn,Demir:2007dt}. For example, a superpotential
operator $W=S L H_u \nu^c/M$ could generate a small Dirac mass in the correct range
for $ \langle S \rangle \sim 10^6$ GeV and $M\sim 10^{18}$ GeV. Such a VEV can easily occur
in intermediate scale models~\cite{Cleaver:1997nj,Langacker:1998ut} or in the \zpr\ mediation scenario~\cite{Langacker:2007ac}. Higher powers  could occur for a larger  $ \langle S \rangle$ or
smaller $M$, associated, e.g., with an anomalous \uprm~\cite{Gogoladze:2001kj}.
Non-holomorphic (wrong Higgs) terms (see Section~\ref{nonholomorphic}) can
also lead to naturally small Dirac masses, suppressed by the ratio of the SUSY breaking
and mediation scales~\cite{Demir:2007dt}. In some cases, a $ Z'$-gaugino is
needed to generate a fermion mass at loop level from a non-holomorphic soft term.
In all of these mechanisms, some low energy
symmetry such as a \uprm\ must forbid a renormalizable level Dirac mass term $W= L H_u \nu^c$,
while allowing the HDO. (The renormalizable level term is allowed in the $E_6$ models.)
Discrete gauge symmetries (i.e., remnants of a gauge symmetry broken at
a high scale), may also help restrict the allowed operators~\cite{Luhn:2007gq}.

Some mechanisms~\cite{Ma:1995xk,Langacker:1998ut} also allow the generation of light sterile neutrino masses and ordinary-sterile
mixing, as suggested by the LSND experiment.

Recently, it was shown~\cite{Nelson:2007yq} that the LSND and MiniBooNE results could
be reconciled in a model involving ordinary and sterile neutrinos if there is a very
light ($\sim$ 10 keV) \zpr\ coupled to $B-L$ with a very weak coupling ($\lesssim 10^{-5}$).
In analogy with the MSW effect~\cite{Mohapatra:2005wg} the \zpr\ generates a potential in matter that is
different for the ordinary and sterile neutrinos and strongly energy dependent.

The right handed components of light Dirac neutrinos could upset the successful predictions
of big bang nucleosynthesis if they were present in sufficient numbers. Mass and
Yukawa coupling effects are too small to be dangerous. However, couplings of the $\nu^c$ to
a TeV-scale \zpr\ could have kept them in equilibrium until relatively late~\cite{Olive:1980wz}.
A detailed estimate~\cite{Barger:2003zh} found that too much $^4He$ would have been produced
for light Dirac neutrinos for most of the $E_6$ models unless $\mzp \gtrsim 1-3$ TeV.
Similar constraints follow from supernova cooling~\cite{Rizzo:1990mx}. 
These limits disappear, however, for couplings close to the $N$ model. This is especially relevant
for  a parameter range of the generalized $E_6$ model (with two \uprm~s) in
which the $Z_N$ is much lighter than the orthogonal boson, but nevertheless no
Majorana masses are allowed~\cite{Kang:2004ix}.

\subsection{Cosmology}\label{sec:cosmology}

\subsubsection{Cold Dark Matter}\label{sec:cdm}
\uprm\ models~\cite{deCarlos:1997yv,Barger:2004bz,Barger:2007nv,Nakamura:2006ht,Lee:2007mt,Hur:2007ur,Belanger:2007dx,Pospelov:2007mp}, 
as well as other singlet extended models with a dynamical $\mu$ term~\cite{Accomando:2006ga,Barger:2007ay,Barger:2007nv,Menon:2004wv}, have many implications for cold dark matter (CDM). For example, the extended neutralino
sector in {Eq.~\ref{eqn:neutmtx}} allows the possibility of a light singlino as the LSP~\cite{deCarlos:1997yv,Barger:2004bz,Barger:2007nv,Nakamura:2006ht,Menon:2004wv},
with efficient annihilation into a light \zpr\ or into the $Z$ (via small admixtures with the Higgsinos).
More generally, the LSP may contain admixtures of $ \tilde S$ or $ \tilde Z'$ with the MSSM
neutralinos, or allow a modified MSSM composition for the LSP. The models also have enlarged
Higgs sectors and different allowed ranges, extending the possible mechanisms for Higgs-mediated
LSP annihilation. Most of the interesting cases should be observable in direct detection experiments~\cite{Barger:2007nv}.

There are other LSP candidates in \uprm\ models. For example, the
scalar partners $\tilde \nu^c$ of the singlet neutrinos become viable thermal CDM candidates due to the
possibility of annihilation through the \zpr~\cite{Lee:2007mt}. Other possibilities include
a neutral exotic particle or multiple stable particles~\cite{Hur:2007ur},
a heavy Dirac neutrino~\cite{Belanger:2007dx}, or a semi-secluded weak sector~\cite{Pospelov:2007mp}
coupled via a \zpr.

\subsubsection{Electroweak Baryogenesis}\label{sec:EWBG}
The seesaw model of neutrino mass allows the possibility of explaining the
observed baryon asymmetry by leptogenesis, i.e., the decays of the heavy Majorana
neutrino generate a small lepton asymmetry, which is partially converted to a baryon
asymmetry by the electroweak sphaleron process~\cite{Mohapatra:2005wg}.
As discussed in Section~\ref{sec:neutrino}, however, an additional \uprm\ symmetry often
forbids the seesaw model. Some of the alternatives discussed there allow other
forms of leptogenesis~\cite{Chun:2005tg}. 

However, the \uprm~\cite{Kang:2004pp,Ham:2007xz,Ham:2007wc} and other singlet extended models~\cite{Barger:2007ay,Menon:2004wv,Profumo:2007wc}
open up the possibility of a completely different mechanism, {\em electroweak baryogenesis}.
In this scenario, the interactions of particles with the expanding bubble wall from a
strongly first order electroweak phase transition lead to a CP asymmetry, which
is then converted to a baryon asymmetry by sphaleron processes.
However, the SM does not have a strong first order transition or sufficient CP violation;
the MSSM has only a small parameter range involving a light stop for the transition, and there is tension between the CP violation needed and electric dipole moment (EDM) constraints~\cite{Carena:2002ss}.
In the extended models, however, there is a tree-level cubic scalar interaction
(the $ \lambda_S A_S S H_u^0 H_d^0$ term in {Eq.~\ref{MSSMUV}}) which can easily
lead to the needed strong first order transition. There are also possible new sources of tree-level CP violation in the Higgs sector~\cite{Kang:2004pp,Ham:2007wc}, which can contribute to the baryon asymmetry but  have negligible effect on EDMs.

\subsubsection{Cosmic Strings}\label{sec:cosmicstrings}
A broken global or gauge \uprm\ can lead to {\em cosmic strings}, which are allowed cosmologically
for a wide range of parameters and which could have interesting
implications for gravitational waves, dark matter, particle emission, and gravitational
lensing. For a recent discussion, with
emphasis on breaking a supersymmetric \uprm\ along an almost flat direction, see~\cite{Cui:2007js}.

\section{Conclusions and Outlook}
A new \uprm\ gauge symmetry is one of the best motivated extensions of the standard model.
For example, \uprm~s occur frequently in superstring constructions.
If there is supersymmetry at the TeV scale, then both the electroweak and \zpr\ scales
are usually set by the scale of soft supersymmetry, so it is natural to expect 
\mzp\ in the TeV range. (One exception is when the \uprm\ breaking occurs along an
approximately flat direction, in which case a large breaking scale could be associated with fermion mass
hierarchies generated by higher-dimensional operators.) 
Similarly, TeV-scale  \uprm~s (or  Kaluza-Klein excitations of the photon and $Z$) frequently occur in models of dynamical symmetry breaking, Little Higgs models, and  models with TeV$^{-1}$-scale extra dimensions.
Other constructions, such as non-supersymmetric grand unified theories larger than $SU(5)$,
also lead to extra \uprm~s, but in these cases there is no particular reason to expect
breaking at the TeV scale (and breaking below the GUT may lead to rapid proton decay).

The observation of a \zpr\ would have consequences far beyond just the existence
of a new gauge boson. Anomaly cancellation would imply the existence of new fermions.
These could just be right-handed neutrinos, but usually there are
additional particles with exotic electroweak quantum numbers. There must also be at least one
new SM singlet scalar whose VEV breaks the \uprm\ symmetry. This scalar could mix with
the Higgs doublet(s) and significantly alter the collider phenomenology. 
The \zpr\ couplings could be
family nonuniversal, allowing new tree-level contributions, e.g.,  to $t$, $b$, and $\tau$ decays.

In the supersymmetric
case the  \uprm\ could solve the $\mu$
problem by replacing $\mu$ by a dynamical variable linked to the \uprm\ breaking, and
the allowed MSSM parameter range would be extended.
The singlets and exotics would be parts of chiral supermultiplets, and there would be extended
neutralino sectors associated with the new singlino and gaugino, modifying the collider physics
and cold dark matter possibilities.  Gauge unification could be maintained if the exotics fell into
$SU(5)$-type multiplets. The \uprm\ symmetry would also constrain the possibilities for neutrino mass and might be related to proton stability and $R$-parity conservation. A \zpr\ might also couple to a hidden sector and could play a role in supersymmetry
breaking or mediation. Finally, a  dynamical $\mu$ would allow a strong first order
electroweak phase transition and new sources of CP violation in the Higgs sector,
making electroweak baryogenesis more likely than in the SM or the MSSM,
with the ingredients observable in the laboratory.

There are large classes of \zpr\ models, distinguished by the chiral charges of the
quarks, leptons, and Higgs fields, as well as the Higgs and exotic spectrum, 
gauge coupling, \zpr\ mass, and possible mass and kinetic mixing.
In string constructions, for example, \uprm~s that do not descend through
$SO(10)$ or left-right symmetry can have seemingly random charges.
 There is no
simple classification or parametrization that takes into account all of the
possibilities. One (model independent) approach, valid for family universality, is to take
a conventional value for the new gauge coupling, and regard the
charges of the left-handed quarks ($Q_L$), leptons ($L_L$),
and antiparticles $u^c_L, d^c_L,$ and $e^+_L$, as well as \mzp, $\Gamma_{Z'}$ and
the mixing angle $\theta$ as free parameters
relevant to experimental searches. However, 8 parameters are too many
for most purposes, so one must resort to specific models or lower-dimensional parametrizations to illustrate the possibilities. A recommended set are those summarized in
Tables \ref{tab:t3rtbl}, \ref{tab:e6decomp}, and \ref{tab:gaugeunif}.

Table~\ref{tab:t3rtbl} lists the $U(1)_{3R} \x U(1)_{BL}$ model, which is a one-parameter (not counting 
the \zpr\ mass and mixing)
set of models based on various forms of $SO(10)$ and left-right symmetry,
and a two parameter generalization motivated by more general embeddings or by kinetic mixing.
It requires no exotics other than $\nu^c_L$. However, the supersymmetric version 
requires non-chiral Higgs doublets and (probably) vector pairs of SM singlets, and does
not solve the $\mu$ problem.

Table~\ref{tab:e6decomp} lists popular $E_6$-motivated models. A whole class
of interesting models involves one free parameter, $\theta_{E_6}$, or
a two parameter generalization with kinetic mixing (or a third parameter if the
gauge coupling is varied). These models illustrate typical exotics, and
(with the exception of the $\chi$ model) the supersymmetric version involves
a dynamical $\mu$ term. However, supersymmetric gauge unification requires
an additional vector pair of Higgs-like doublets.

The models in Table~\ref{tab:gaugeunif} are examples of supersymmetric models
with a dynamical $\mu$ that are consistent with gauge unification without additional
vector pairs. Three parameters, including the gauge coupling, are relevant to
the non-exotic sector.

If there is a  \zpr\ with typical electroweak scale couplings to the
ordinary fermions, it should be readily observable at the LHC for masses up to $\sim 4-5$ TeV, or at the Tevatron for masses up to $\sim 600-900$ GeV. Significant diagnostic probes of the
\zpr\ couplings would be possible up to $\sim 2-2.5$ TeV. A future ILC would extend the
range somewhat, and would provide complementary diagnostics. 
Within the context of supersymmetry, the observation of a \zpr\ could completely alter the paradigm of having
just the MSSM at the TeV scale, with a desert up to a scale of grand unification or
heavy Majorana neutrino masses, and would suggest a whole range of new 
laboratory and cosmological consequences. In the nonsupersymmetric case,
a \zpr\ might be one of the first experimental manifestations of a new TeV scale
sector of physics.

\section*{Acknowledgments}
I am extremely grateful to Vernon Barger, Mirjam Cveti\v c, Jens Erler, and all of
my other collaborators on work related to this article, and Hye-Sung Lee for a careful reading of the
manuscript. This work was supported by the
Friends of the IAS and by NSF grant PHT-0503584.

\bibliography{zpref}
\end{document}